\documentclass[11pt,twosided]{article} 

\usepackage{amssymb}
\usepackage{amsfonts}
\usepackage{amsmath}
\usepackage{amsthm}
\usepackage{graphicx}
\usepackage{caption}
\usepackage[usenames, dvipsnames]{xcolor}
\usepackage{verbatim}
\usepackage{dsfont}
\usepackage{xcolor}
\usepackage[utf8]{inputenc}
\usepackage{hyperref}
\usepackage{subfigure}
\usepackage{comment}
 \usepackage{tikz}
\usetikzlibrary{positioning,chains,fit,shapes,calc}

\usepackage{fullpage}
\usepackage{setspace}
\numberwithin{equation}{section}

\newtheorem{theorem}{Theorem}[section]
\newtheorem{lemma}[theorem]{Lemma}
\newtheorem{remark}[theorem]{Remark}
\newtheorem{example}[theorem]{Example}
\newtheorem{proposition}[theorem]{Proposition}
\newtheorem{definition}[theorem]{Definition}

\newtheorem{corollary}[theorem]{Corollary}
\newtheorem{fig}[theorem]{Figure}

\newcommand{\bthe}{\begin{theorem}}
\newcommand{\ethe}{\end{theorem}}

\newcommand{\ben}{\begin{enumerate}}
\newcommand{\een}{\end{enumerate}}

\newcommand{\bit}{\begin{itemize}}
\newcommand{\eit}{\end{itemize}}

\newcommand{\beq}{\begin{equation}}
\newcommand{\eeq}{\end{equation}}

\newcommand{\ble}{\begin{lemma}}
\newcommand{\ele}{\end{lemma}}

\newcommand{\bde}{\begin{definition}\rm}
\newcommand{\ede}{\Chalmos\end{definition}}

\newcommand{\bco}{\begin{corollary}}
\newcommand{\eco}{\end{corollary}}

\newcommand{\bpr}{\begin{proposition}}
\newcommand{\epr}{\end{proposition}}

\newcommand{\brem}{\begin{remark}\rm}
\newcommand{\erem}{\Chalmos\end{remark}}

\newcommand{\bexam}{\begin{example}\rm}
\newcommand{\eexam}{\end{example}}

\newcommand{\bfi}{\begin{fig}}
\newcommand{\efi}{\end{fig}}

\newcommand{\btab}{\begin{tab}}
\newcommand{\etab}{\end{tab}}

\newcommand{\beao}{\begin{eqnarray*}}
\newcommand{\eeao}{\end{eqnarray*}\noindent}

\newcommand{\beam}{\begin{eqnarray}}
\newcommand{\eeam}{\end{eqnarray}\noindent}

\newcommand{\overliner}{\begin{array}}
\newcommand{\earr}{\end{array}}

\newcommand{\bdis}{\begin{displaymath}}
\newcommand{\edis}{\end{displaymath}\noindent}

\def\N{{\mathbb N}}

\def\S{{\mathbb S}}
\def\P{{\mathbb P}}
\def\E{{\mathbb E}}
\def\R{{\mathbb R}}

\newcommand{\pr}[1]{\P\left(#1\right)}

\def\calb{{\mathcal{B}}}

\def\calr{{\mathcal{R}}}

\def\1{\mathds{1}}
\newcommand{\bone}{{1}}
\newcommand{\bnull}{{0}}

\newcommand{\stv}{\stackrel{v}{\rightarrow}}

\newcommand{\tto}{{t\to\infty}}

\newcommand{\nto}{{n\to\infty}}

\newcommand{\al}{{\alpha}}

\newcommand{\si}{{\sigma}}

\newcommand{\supp}{{\rm supp}}

\newcommand{\VaR}{{\rm VaR}}

\newcommand{\CoTE}{{\rm CoTE}}

\newcommand{\dep}{{\rm dep}}
\newcommand{\ind}{{\rm ind}}

\newcommand{\ov}{\overline}

\newcommand{\wt}{\widetilde}

\newcommand{\Chalmos}{\quad\hfill\mbox{$\Box$}}  

\allowdisplaybreaks[4]

\newcommand{\CK}[1]{{\color{black} #1}}
\newcommand{\OK}[1]{{\color{black} #1}} 
\newcommand{\Ored}[1]{{\color{black} #1}}  

\newcommand{\GR}[1]{{\color{black} #1}} 
\newcommand{\gr}[1]{{\color{black} #1}} 
\newcommand{\gdr}[1]{{\color{black} #1}}  
\newcommand{\Gr}[1]{{\color{black} #1}} 

\begin{document}

\vspace{-3cm}
\title{Risk in a large claims insurance market with bipartite graph structure} 

\author{Oliver Kley\thanks{Center for Mathematical Sciences, Technische Universit\"at M\"unchen,  85748 Garching, Boltzmannstrasse 3, Germany, e-mail: oliver.kley\,,\,cklu@tum.de}
\and Claudia Kl\"uppelberg\footnotemark[1]
\and Gesine Reinert\thanks{Department of Statistics, University of Oxford, 1 South Parks Road, Oxford OX1 3TG, UK, email: reinert@stats.ox.ac.uk }
}

\maketitle

\abstract{%
{We  model the influence of sharing large exogeneous losses to the reinsurance market  by a bipartite graph.} Using Pareto-tailed claims and multivariate regular variation we obtain asymptotic results for the Value-at-Risk and the Conditional Tail Expectation. 
We show that the dependence on the network structure plays a fundamental role in their asymptotic behaviour. 
As is well-known \gdr{in a non-network setting}, if the Pareto exponent is larger than 1,  then for the individual agent (reinsurance company)  diversification is beneficial, whereas when it is less than 1, concentration on a few objects is the better strategy. 

\GR{An additional aspect of this paper is the amount of uninsured losses which have to be convered by society. In the situation of networks of agents, \gdr{in our setting} 
 diversification is never detrimental concerning the amount of uninsured losses.  If the Pareto-tailed claims have finite mean, diversification turns out to be} \gdr{never detrimental}, both for society and for individual agents. \gdr{In contrast}, \GR{if the Pareto-tailed claims  \gdr{have infinite} mean, a conflicting situation may arise between the incentives of individual agents and the interest of some regulator to keep risk for society small.
We  explain the influence of the network structure on diversification effects  in different network scenarios.}

\noindent
\begin{tabbing}
{\em MSC2010 Subject Classifications:} \= primary:\,\,\,90B15\,\,\,
secondary: \,\,\,60G70, 62P05, 91B30, 62E20 \\
\end{tabbing}

\section{Introduction}\label{s1}

Over the last years, risk modelling has increasingly taken the fact into account that agents are related through an interwoven network of business relationships, see \cite{Bossetal,BrunnCher,CMS, EN, GaiKap, haldane2011systemic} to mention just a few among a rapidly increasing number of articles on the topic. 
\CK{For a financial market the notion systemic risk is often applied to model endogeneous risk in a (banking) network leading to cascading behaviour. 

Insurance risk is of a different flavour, since exogeneous risks play the essential role, although the market system covering jointly these risks, may also have some influence (cf. \cite{bain1997insurance, lin2014reinsurance}).
Yet, the specific nature of the extremal dependence structure between large losses, as they may for example happen in markets for catastrophe insurance, has not yet been taken into account in such models. This paper contributes to filling this gap and we understand risk here as the risk to the proper functioning of the system; see \cite{zigrand}. We refer to this not only by studying the role of risk sharing for reinsurance agents in different market situations but also by evaluating the extent of large losses which are not covered by the reinsurance industry and hence remain to the society, cf. \cite{GvPuninsured}. A main feature of insurance risks is that they are heavy-tailed, while often obeying a law of regular variation.}

\CK{Regular variation is a powerful tool which can and has been used in many areas of applied probability. In particular, 
it is a standard concept in insurance risk models with focus on the ruin probability as a risk measure. Whereas one-dimensional risk processes have been studied since Cram\'er introduced the compound Poisson insurance risk processes in the 1930s, ruin problems for multivariate models are somewhat scattered in the literature. 
For instance, \cite{HLunpub} study ruin problems for an insurance company with multiple business lines allowing for capital transfer between these lines.
They investigate so-called ruin regions; i.e. some far out set in $\R^d$ such that the multivariate risk process hits (for some time $t>0$ or some time $t\in (0,T)$) this set with very small probability.
This leads to some rules for risk and capital transfer to possibly avoid insolvency.
Similar problems of optimal reserve allocation have also been considered in this context; see for example \cite{Biard}.
The leading model in these papers are multivariate compound Poisson processes with some rather restricted dependence structure between the claims in the different business lines (cf. Section 4 of \cite{HLunpub}). 
The ruin problem in a linear portfolio of insurance risk processes has also been investigated in \cite{BKruin}, where the risk processes may again represent business lines or different companies.
Dependence between the risk processes is modeled by a Clayton dependence structure, which allows for scenarios reaching from weak to very strong dependence.
Much more general L\'evy models have been investigated in \cite{EK1,EK2}, where the probabilistic sample path behaviour leading to ruin is described.
Risk assessment in a class of networks with two types of participants, insurance and reinsurance companies, is modelled with the goal of simulating scenarios in \cite{BlanchetShi}. 
It is assumed that the claim sizes have a linear factor model structure with Pareto-tailed factors.
Assumptions are further that contractual relationships remain constant, when there is no default, and reduce to a subgraph, if default occurs. The authors adopt an equilibrium approach closely related to the market clearing framework established in \cite{EN}. }


\CK{
Moreover, a famous result in graph theory by \cite{bollobas} was recently extended in \cite{SRTDWW} and ensures that in a preferential attachment model the in- and out-degrees are multivariate regularly varying. 
For such power-law financial networks \cite{AminiMinca} give conditions under which a higher order cascade dies out.}

Our results build on this literature and extend it to a \CK{market setting. The main aim of this paper is to assess the effect of networks on high losses under general assumptions on the insurance risks.}
Our model assumes a finite set of agents (reinsurance companies), and a finite set of objects.  We think of each object as a pool of highly dependent risks which cause a severe loss if  one common triggering event happens. To give examples, we could think of object $1$ as a portfolio of household insurances in a particular hurricane region in the U.S. Object 2 could be a pool of life insurances in Western Europe which might cause a { large}  claim in case an epidemic happens, and Object 3 might be related to an earthquake in Japan.   Often, we consider the object claims to be asymptotically stochastically independent, which is not a severe restriction of generality if they either refer to different sources of risk (hurricane vs. fire) or to different geographical regions (earthquake in Japan vs. earthquake in California). \Ored{If an object is chosen by some agent then it is at least partly insured.} 
In a realisation of the network, some  objects  may not be  insured, which happens also in reality, e.g. in less developed 
regions 
of the world. An object generates a loss with a Pareto-tail if it gets severely damaged. \OK{Then the loss is  distributed across all agents insuring that object.}

Hence we use a bipartite graph as in  Figure~\ref{fig1} to model  the  structure of the reinsurance market; this model strongly resembles the depiction of the reinsurance market in
Figure~21 of \cite{IAIS}.

\begin{figure}[t]
\vspace*{-2cm}
\begin{center}
\includegraphics[width=0.9\textwidth]{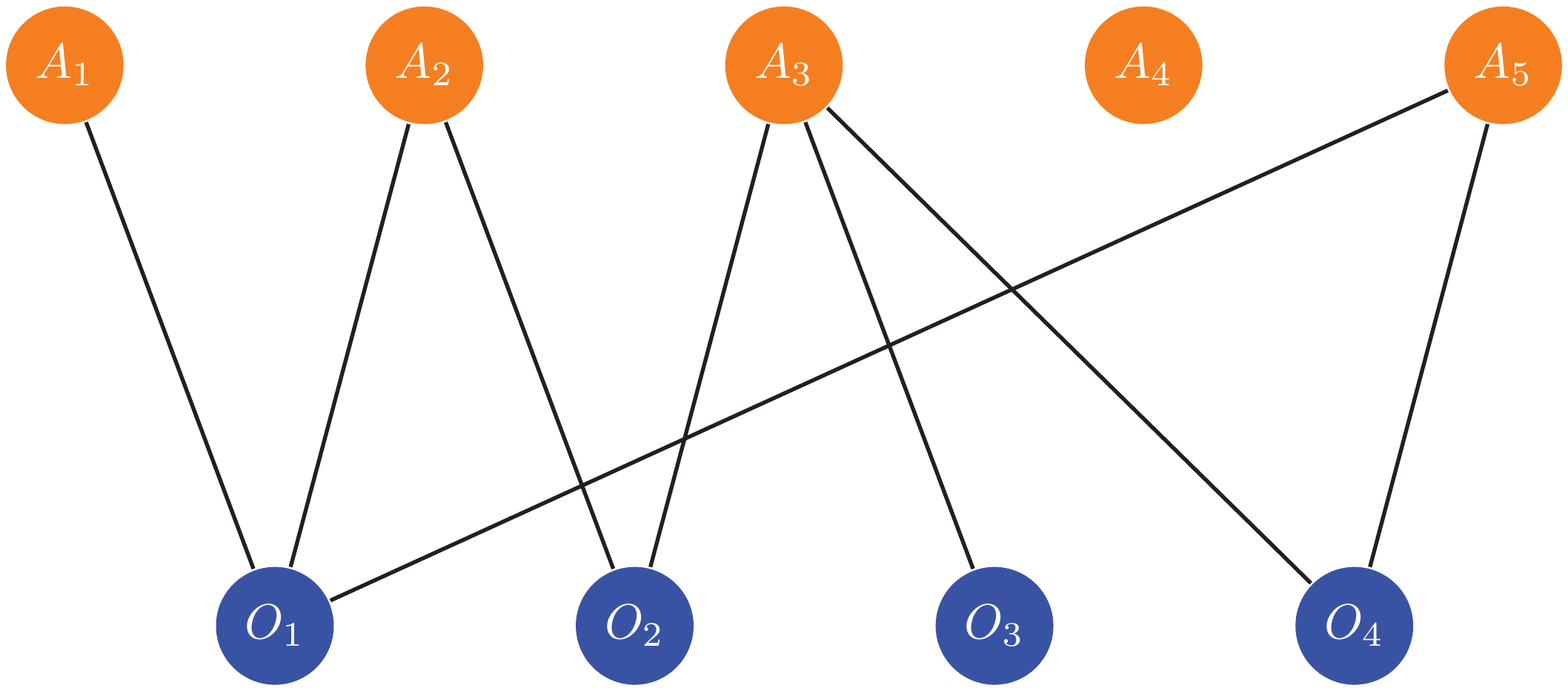}
\vspace*{-2cm}
\end{center}
\caption{\label{fig1} The hierarchical structure of the reinsurance market as a {bipartite graph}.}
\end{figure} 

While  our underlying structure is a bipartite graph of agent-object relationships, we mainly focus on the agents and their exposure in the analysis \Ored{ as well as the amount of uninsured risk which has to be covered by society}.

This model is intentionally very simple and does not attempt to capture the minutiae of the reinsurance market. In particular the model does not reflect the fact that reinsurers may reinsure one another, potentially leading to spiralling effects, see for example \cite{bain1997insurance}. The reason to exclude this effect is that retrocession nowadays covers just a small fraction of the overall reinsured risk, see \cite{IAIS}, p. 26. Instead our model  focuses on the effect of the network between agents and objects on risk assessment. {{This paper could serve as a starting point for further investigations including balance sheets and address risk allocation.}} 

In this paper we address the following questions: Taking network effects into account, when is it beneficial for an agent to diversify, and when is it beneficial to concentrate on a few objects? From a macroscopic viewpoint, when is it beneficial to have  highly diversified agents, and when would it be beneficial that agents focus only on a few objects instead? Here the benefit is judged according to  the Value-at-Risk and  Conditional Tail Expectation
of  individual agents as well as the whole system which allow us to take the macro-prudential and the micro-prudential view.

The use of bipartite networks has been successful in the area of common asset holdings, see for example \cite{braverman2014networks, Cacciolietal}. 
While \cite{braverman2014networks} assesses price impacts on assets due to shocks to other assets, \cite{Cacciolietal} uses a stylized mechanism to quantify bankruptcy cascades.
We will instead concentrate on {determining} extremal dependence structures of losses with  Pareto-tails using the framework of multivariate regular variation. 
Note that regular variation techniques have already been applied  to the problem of portfolio diversification with respect to heavy-tailed risks; see for example~\cite{Ibragimov2005, MR, Zhou}. 
Our results build on this literature and extend it to a {{network}} setup. 
While previous research considers standard portfolios of different risk factors without any network structure, we model the whole market of  various portfolios and describe their dependence structure, which is  determined by the bipartite random graph. 

In more detail, our \textit{random bipartite  graph} is constructed as follows.
We have a vertex set of agents  $\mathcal{A}$ of cardinality $q$ as well
as a  vertex set of objects $\mathcal{O}$ of cardinality $d$.
Each vertex (agent) $i \in \mathcal{A}$ chooses a number of objects from $\mathcal{O}$ to which it is linked, where each object $j\in \mathcal{O}$ is chosen independently with probability $p_{ij}\in[0,1]$ (thus including the case of a deterministic graph).
Different agents choose their object sets independently from each other.
Let $V_{j}$ for $j=1,\dots,d$ have Pareto-tails  determining the claim, if object $j$ gets severely damaged, so that, for possibly different $K_j>0$ and tail index $\al > 0$,
\beam\label{pareto}
P(V_j>t)\sim K_j t^{-\al},\quad t \to \infty.
\eeam
(For two functions $f$ and $g$ we write $f(t)\sim g(t)$ as $\tto$ if $\lim_{\tto} f(t)/g(t) = 1$.)
We summarize all claims in the vector $V=(V_1,\dots,V_d)^\top$ and assume that this vector
is independent of the random graph construction, while $V_1,\ldots,V_d$ may not be independent of each other.
Obviously, losses may have different tail indices $\alpha$. 
Although to require the same $\alpha$ seems restrictive, indeed it is not. 
Two remarks are in order:
\begin{enumerate}
\item[(i)] If loss categories have different $\alpha$'s, then the famous Breiman result \cite{Breiman} ensures that
the aggregated losses have a tail behaviour corresponding to the smallest $\alpha$: the largest losses dominate.
As a prerequisite of our analysis we would estimate $\alpha$ for each object and a statistical test would lead to the relevant subclass of objects to be analysed. 
\item[(ii)] On the other hand, if we want to include also the less severe losses into our analysis, we can unify all marginal tails for instance to the smallest $\al$ (similar as for copulas) and base our analysis on the full sample (cf. Proposition~5.10 of \cite{Resnick1987}.).
\end{enumerate}

Let $f_{i}(V_{j})$ now be the proportion of the loss of object $j$ which affects agent $i$. 
\OK{We assume that the loss is distributed 
across all agents insuring object $j$, and we indicate that agent $i$ holds some insurance risk on object $j$ by $\mathds{1}( i\sim j)$.
Then
$$
f_{i}(V_j)={\1(i\sim j)} W_{ij} V_j.
$$
The weights \gdr{$W_{ij}\ge 0$}
 indicate the proportion of claim $j$ agent $i$ ensures and may also be random, \gdr{ and could be} coupled to the random network structure. The sum $\sum_{i=1}^{q}W_{ij}\1(i\sim j)$ gives the total part of object $j$ which is insured.} 
Let $F_i:=\sum_{j=1}^df_{i}(V_j)$ denote the {\em exposure of agent $i$} and 
$F=(F_1,\dots,F_q)^\top.$
We represent the insurance relations through the weighted  $q\times d$ adjacency matrix  
 $A=(A_{ij})_{i,j=1}^{q,d}$  with \OK{
 \begin{gather}\label{eq2.2}
 A_{ij} = \1 (i\sim j) W_{ij}.
 \end{gather}
We detail some specific cases of the weights $W_{ij}$ after Theorem~\ref{singleFasym}. }
As a consequence of \eqref{eq2.2} we can represent the vector of exposures of the $q$ agents as
\begin{gather}\label{F}
F=AV.
\end{gather}

Our model is related to the reinsurance network model in \cite{lin2014reinsurance}, where the relationship between primary insurers and reinsurers is modelled instead of the relationship between objects and reinsurers; there, an insurer's risky asset is modelled as normally distributed. The crucial difference is that our model includes  the heavy-tailed nature of the losses, which requires a {very}  different treatment. 

The Pareto-tails allow us to assess the exposure of an agent, the vector of exposures of all agents as well as the aggregated exposures, defined by the norm of the exposure vector.
 Here and throughout the paper $\|\cdot\|$ is a norm in $\R^d$ or $\R^q$ such that \OK{all canonical unit vectors have norm 1.}
We will show at the end of Section~\ref{s3}
that the effect of the network structure on the Value-at-Risk and the  Conditional Tail Expectation,  given the tail index $\alpha$ of the Pareto-tailed claims, is solely contained in the  constants 
\begin{equation} \label{VaRconst}
 C^i_{ind} =  C^i_{ind} (A) := \sum_{j=1}^{d} K_j\E \ A_{ij}^{\alpha},  \,\, i=1, \ldots, q,  \quad \mbox{ and } \quad C_{ind}^S = C_{ind}^S (A) = \sum_{j=1}^d K_j \E  \ \| A e_j \|^\alpha , 
\end{equation}
when the Pareto-tailed claims are independent; the superscript $i$ indicates the individual setting of agent $i$, whereas $S$ refers to the systemic setting {{in the sense that the system of all agents and objects is taken  into account}}. 
We contrast this with the fully dependent case. Defining the $d\times d $ diagonal matrix $K^{1/\al}={\rm diag}(K_1^{1/\al},\ldots,K_d^{1/\al})$, the corresponding quantities are 
\begin{gather}\label{VaRconst_dep}   
C_{dep}^i = C_{dep}^i (A) :=  \E (AK^{1/\al}\bone)_{i}^{\alpha },  \,\, i=1, \ldots, q,   \quad \mbox{ and }\quad  C_{dep}^S = C_{dep}^S (A) = \E \|AK^{1/\al}\bone\|^{\alpha},
\end{gather}
where $\bone$ is the $d-$dimensional vector with entries all equal 1.
In general, small constants are more desirable, indicating a smaller risk.
The case of fully dependent claims is equivalent to having a single source of loss, but the loss to be unevenly distributed among the agents. 

As \cite{Ibragimov2005, MR, Zhou}, we find that for every individual agent  diversification is beneficial, whereas in the case that $\alpha < 1$ \gdr{so that the risks have infinite mean}, concentration on a few objects is the better strategy. 
\GR{ From the perspective of a society, however,  uninsured losses \gdr{are not} desirable. \gdr{The  better connected and consequently more diversified markets are, the fewer uninsured losses would be expected}. Thus, if  $\alpha < 1$ there is a conflict between an agent's wish  not to diversify  and the society's wish to have all of the objects insured. If $\al>1$ \gdr{then}  the individual agent's wishes and the society's need for risks \gdr{to be covered agree}. 
}

To disentangle the different factors which enter into the risk calculations, \GR{ namely the dependence information and the degree of heavy-tailedness of exogeneous losses as well as the network structure of the reinsurers with the objects,} our paper is organised as follows. Section \ref{ss22} collates results from multivariate regular variation. In Section \ref{s3} we review and derive specific asymptotic results for the regular variation of $F$ under a general dependence structure for $V$, and then we consider the special cases that $V$ has asymptotically independent components, and that $V$ has asymptotically fully dependent components. Here the (random) network structure enters as a (stochastic) linear transformation of the claims vector $V$.
We apply our results to the risk measures Value-at-Risk and Conditional Tail Expectation. \GR{This section also states bounds for general dependence structures.
Section \ref{s53}  discusses the effect of the network on the risk measures and gives some examples for bipartite network models.  We prove Poisson approximations of relevant constants for a large market.} \gdr{While most of the examples in this section use the matrix $A$ with 
\begin{gather}\label{eq2.21}
 A_{ij} = \frac{\1 (i\sim j)}{\deg(j)} \gdr{, \quad \mbox{ where } \frac{0}{0}:=0,}
\end{gather}
and $\| \cdot \|$ the 1-norm, we conclude the section with 
a discussion of the use of an arbitrary $r-$norm with  $r>1$ or $r-$quasinorm with $r<1$ as aggregation function for the risk in the market.
}

\section{Preliminaries from multivariate regular variation}\label{ss22}

Our framework is that of regular variation of the random vector of exposures $F$, which follows from the Pareto-tailed claims and the dependence structure introduced by the bipartite graph.
We start with a set of equivalent definitions; cf. Theorem~6.1 of \cite{Resnick2007}, and Ch.~2.1 of \cite{BasrakPhD}.
For $d\in\N$,  $\mathbb{S}_{+}^{d-1}=\{ x\in \R^{d}_{+}: \ \| x\|=1 \}$ is the positive part of the  unit sphere in $\R^{d}$ with respect to the norm $\|\cdot\|$.
We denote by $\bnull$ the $d-$dimensional vector with entries all equal to 0, and write for short 
$$\mathcal{E}:=\overline\R_+^d\setminus{\{\bnull\}}$$ 
 with $\overline{\R}_+=[0,\infty]$.  Moreover, we denote by  $\calb=\calb(\mathcal{E})$ the Borel $\si$-algebra with respect to  the so-called \textit{one point uncompactification}, cf.~\cite{Resnick2007}, Section~6.1.3. In this topology, the compact sets are exactly the sets which are compact in $\overline\R_+^d$ and which do not contain the zero vector $\bnull$. In particular, sets of the form $(x,\infty]$ for $x \in \overline{\R}_+^{d}\setminus\{\bnull\}$ are relatively compact and bounded sets in this topology are the sets which are bounded away from the origin.
We say that a sequence of Radon measures $(\nu_{n})_{n \in \N}$ on $\mathcal{B}$ \textit{converges vaguely}  to a Radon measure $\nu$ on $\mathcal{B}$, for short $\nu_n \stackrel{v}{\to} \nu$, if 
$
\int_{\mathcal{E}}f d\nu_{n} \rightarrow \int_{\mathcal{E}}f d\nu$ as $n \to \infty$
for all measurable functions $f$ on $\mathcal{E}$ which are continuous with compact  support. For equivalent definitions of vague convergence see Proposition~3.12 in \cite{Resnick1987}.

\bde\label{RV}
A random vector $X$ with state space $\mathcal{E}$
is called {\em multivariate regularly varying} if one of the following equivalent conditions holds:\\
(i) \, There is a Radon measure   $\mu\not\equiv 0$ on $\mathcal{B}$ with $\mu(\overline\R_+^d\setminus\R^d_+)=0$ and 
 a relatively compact Borel set $D \in\mathcal{B}$ such that $tD$ 
is bounded away from zero and $\mu(\partial tD)=0$ for $t$ in a dense subset $T$ of $(0,\infty)$ and 
\begin{gather}\label{basrakmu} 
\frac{\P(X\in t\cdot)}{\P(X\in t D)}\stv \mu(\cdot),\quad\tto.\end{gather}
In that case there exists some $\al>0$ such that the limit measure is homogeneous of order $-\al$:
$$\mu(u S)=u^{-\al}\mu(S)$$ 
for every  $S\in\calb$ satisfying $\mu(\partial S)=0$. The measure $\mu$ is called \textit{intensity measure of $X.$} 
Note that $\mu$ has no atoms. \\[2mm]
(ii) \, There is a Radon measure $\nu\not\equiv 0$ on $\mathcal{B}$  
 and a sequence $b_{n}\rightarrow \infty$
so that
\begin{gather}\label{defMVR}
n\P\Big(\frac{X}{b_{n}}\in \cdot\Big)\stackrel{v}{\rightarrow} \nu(\cdot),\quad n \rightarrow \infty.
\end{gather} 
The measure $\nu$ is homogeneous of the  same order $\alpha$ as $\mu$  in (i), and   is called the {\em exponent measure of $X$}.
\\[2mm]
(iii) \,  There is a probability measure $\rho\not\equiv 0$ on $\calb (\S_+^{d-1})$ such that for some $\al > 0$
\beam\label{RV1}
\frac{\P(\|X\|>tu, X/\|X\|\in\cdot)}{\P(\|X\|>t)}\stv u^{-\al} \rho(\cdot),\quad\tto,
\eeam
for every $u>0$. The index $\al$ is the same as in (i) and (ii).
The measure $\rho$ is called the {\em spectral measure of $X$}.\\[2mm]
The {\em  tail index} $\al > 0$ is also called the  index of regular variation of $X$, and we write $X \in \mathcal{R}(-\alpha).$
\ede

There is a certain choice in the normalization of the left hand sides of \eqref{basrakmu} (the choice of $D$) and \eqref{defMVR} (the choice of $b_n$). 
As an example consider the one-dimensional case of a Pareto-tail as in \eqref{pareto}.
If we choose $b_n\sim (K_1 n)^{1/\al}$, then 
$$n\pr{V_1>(K_1 n)^{1/\al} t }\sim n (K_1 n)^{-1} K_1 t^{-\al} =  t^{-\al},\quad\nto.$$
Alternatively, we can choose $b_n\sim  n^{1/\al}$ and obtain
\beam\label{normalization}
n\pr{V_1>n^{1/\al} t }\sim  K_1 t^{-\al},\quad\nto.
\eeam
We will use the second normalization, applying the same $b_n$ to each vector component of $V$, and retain the different constants $K_j$ in the limit. 

The measures $\mu$, $\nu$ and $\rho$ all assess the dependence structure of the multivariate vector $V$ in the limit. 
If $\mu$, $\nu$ and $\rho$ have only mass on the axes, then they correspond to independence, and we say the components of $V$ are asymptotically independent. 
If the mass is concentrated only on the line $\{s K^{1/\al}\bone: s > 0\}$, then $\mu$, $\nu$ and $\rho$ correspond to full dependence, and we say $V$ is asymptotically fully dependent. 
We present the corresponding measures $\nu$ and $\mu$ in Lemmas~\ref{lemma:depindep} and~\ref{lemma:indep} below; cf. Section~6.5.1 in~\cite{Resnick2007}.


\begin{lemma}\label{lemma:depindep}
Assume that the vector $V:=(V_1,\ldots,V_d)^\top$, whose components have Pareto-tails as in \eqref{pareto}, belongs to $\calr(-\al)$. 
Define $b_n:= n^{1/\al}$ and let $x=(x_1,\ldots,x_d)^\top$. Then the exponent measure $\nu$  from \eqref{defMVR} of $V$ is given by
\begin{enumerate}
\item[(a)]  \label{nuindependent}  $\nu([0,x]^{c})= \sum_{j=1}^{d} K_j x_{j}^{-\alpha} $ if the components of $V$ are asymptotically independent, and 
\item[(b)] \label{nudependent}   $ \nu([0,x]^{c})= \max_{j=1,\dots,{d}} \{K_j x_{j}^{-\alpha}\} $ if  $V$ is asymptotically fully dependent.
\end{enumerate}
\end{lemma}

\proof{Proof:} \,  
Define $V^*:=(K_1^{-1/\al}V_1,\ldots,K_d^{-1/\al}V_d)^\top$, then all component have the same Pareto-tail
$P(K_j^{-1/\al}V_j>t)\sim t^{-\al}$ as $\tto$ for $j=1,\dots,d$. If we denote by $\nu^*$ the exponent measure of $V^*$, then by (6.31) and (6.32) of \cite{Resnick2007} $\nu^*([0,x]^{c})= \sum_{i=1}^{d}x_{i}^{-\alpha} $ if the components of $V^*$ are asymptotically independent, and  $\nu^*([0,x]^{c})= \big(\min_{i=1,\dots,{d}} x_{i}\big)^{-\alpha} $ if  $V^*$ is asymptotically fully dependent.
To obtain the exponent measure of $V$ we summarize the different coefficients in the diagonal matrix
$K^{1/\al}={\rm diag}(K_1^{1/\al},\ldots,K_d^{1/\al})$.
Then $\nu=\nu^*\circ K^{-1/\al}$ and we obtain 
$$\nu([0,x]^{c}) = \nu^*\circ K^{-1/\al}([0,x]^{c}) =\nu^*([0,K^{-1/\al}x]^{c}),$$ 
giving (a) and (b) above.

\endproof

\begin{lemma} \label{lemma:indep}
{Assume that the vector $V:=(V_1,\ldots,V_d)^\top$, whose components have Pareto-tails as in \eqref{pareto}, belongs to $\calr(-\al)$. } 
Consider the intensity measure $\mu$ from \eqref{basrakmu} with $D=\{x\in\overline{\R}^d_+: \|x\|>1\}=\{\|x\|>1\}$ and the exponent measure $\nu$ as in Lemma~\ref{lemma:depindep}. Then
\begin{enumerate}
\item[(a)]
$\mu(\{\|x\|>1\})  = 1$ and
 $\mu(\cdot)=\dfrac{\nu(\cdot)}{\nu(\{\|x\|>1\})}$. 
 \item[(b)]
 $\pr{\|V\|>t}\sim \nu(\{\|x\|>1\}) t^{-\al}$ as $\tto$.
\item[(c)]
$\nu(\{\|x\|>1\}) = \sum_{j=1}^d K_j$, if the components of $V$ are asymptotically independent.
\item[(d)]
$\nu(\{\|x\|>1\}) = \|K^{1/\al}\bone\|^\al$, if the components of $V$ are fully dependent.
\end{enumerate}
\end{lemma}

\proof{Proof:}
(a)  From \eqref{basrakmu} it follows that $\mu(D)=1$.
The convergence $ n \pr{b_{n}^{-1}{V}\in \cdot} \stackrel{v}{\rightarrow} \nu(\cdot)
$ of \eqref{defMVR}  with $b_{n}=b(n)=n^{1/\al}$ yields with continuous parameter $t$
\begin{gather*}
\lim_{t\rightarrow \infty } t^\al\pr{V \in t [0,x]^{c}} = \nu([0,x]^{c}), 
\end{gather*} 
see \cite{Resnick2007}, p.~175.
Consequently, 
\begin{gather}\label{supdings}
\lim_{t\rightarrow \infty}\frac{t^\al\pr{V \in t [0,x]^{c}}}{\pr{V \in t[\bnull,\bone]^{c}}t^\al} = \frac{\nu([0,x]^{c})}{\nu([\bnull,\bone]^{c})}.
\end{gather}
Additionally, employing Lemma~6.1 from \cite{Resnick2007} and noting the relative compactness of $\{\|x\|>1\}$, 
\begin{gather}\label{supdings2}
\lim_{t\rightarrow \infty}\frac{\pr{\|V\|>t }}{\pr{V \in t[\bnull,\bone]^{c}}} = \frac{\nu(\{\|x\|>1\})}{\nu([\bnull,\bone]^{c})}.
\end{gather}
Putting relations \eqref{supdings} and \eqref{supdings2} together leads for fixed $x_0 \neq 0$ to
\begin{gather*}
\frac{\nu([0,x_0]^{c})}{\nu([\bnull,\bone]^{c})}=\lim_{t\rightarrow \infty}\frac{\pr{V\in t [0,x_0]^{c}}}{\pr{\|V\|>t}}\frac{\pr{\|V\|>t}}{\pr{V \in t[\bnull,\bone]^{c}}}= \mu([0,x_0]^{c})\frac{\nu(\{\|x\|>1\})}{\nu(\{[\bnull,\bone]^{c}\})},
\end{gather*}
which gives $(a)$ and $(b)$. \\
In order to prove (c) and (d) we have to calculate the specific form of  $\nu(\{\|x\|>1\})$ for the vector $V$.
Denote by $\nu^*$ the exponent measure of $(K_1^{-1/\al} V_1,\ldots,K_d^{-1/\al} V_d)^\top$, and recall the matrix $K^{1/\al}={\rm diag}(K_1^{1/\al},\ldots,K_d^{1/\al})$.
Then for the asymptotically independent case we calculate
\beao
\nu(\{\|x\|>1\}) &=& \nu^*\circ K^{-1/\al} (\{x\in\R^d : \|x\|>1\})\\
&=& \nu^*(\{x\in\R^d : \|K^{1/\al}x\|>1\})\\
&=& \sum_{j=1}^{d} \nu^*\big(\{t e_j\in\R^d : t > K_j^{-1/\al}\}\big) = \sum_{j=1}^{d} K_j,
\eeao
and for the asymptotically fully dependent case we obtain with \eqref{pareto}
\beao
\nu(\{\|x\|>1\}) &=& \nu^* (\{ t \bone : t\|K^{1/\al}\bone\| >1\}) \, = \, \|K^{1/\al}\bone\|^\al.
\eeao

\endproof

\section{Regular variation of $F$}\label{s3}

We assume the vector $V$ of claims to be multivariate regularly varying with tail index $\alpha>0$, which is for example trivially the case if the marginals have Pareto-tails with tail index $\alpha$ (as in \eqref{pareto}) and are either asymptotically independent or fully dependent.
We now turn to the multivariate vector $F=AV$ of exposures of all agents. In this section, we determine  the dependence measures of $F$, the asymptotics of probabilities of joint marginal exceedances as well as the asymptotics of exceedances of the aggregated vector $F$ in form of a norm of $F$,  all in dependence of the law of the random matrix $A$ representing the market structure. We then apply our findings to Value-at-Risk and Conditional Tail Expectations as tools for risk management.

\subsection{Dependence measures of $F$ and asymptotics}

We start with some notation.
Recall that $\|\cdot\|$ denotes a norm on $\R^d$  or $\R^q$  satisfying $\|e_j\|=1$ for all unit vectors.
We {abbreviate} the cone
$$C_{1,\S_+^{q-1}}:=\{x\in \overline\R^{q}_{+}:\ \|x\|>1 \}.$$
The {\em support} of the random matrix $A$ is
$$\operatorname{supp}(A):=\{ M \in \R^{q\times d}:\ \pr{A=M}>0\}.$$
Further, for $x\in \R^{q}$  the {\em Hamming  distance}  between $x$ and $\bnull$ is given by  
$$|x|_{H}:=|\{i \in \{1,\dots,q \}:\ x_{i}\neq 0\}|.$$
The following result, based on Proposition~A.1 of \cite{Basrak200295},  ensures regular variation of $F=AV$. 
 Note that we use a different normalization compared to \cite{Basrak200295}.
 
\begin{proposition}\label{th1}
Let $V:=(V_1,\ldots,V_d)^\top$ have components with Pareto-tails $\P (V_j>t) \sim K_j t^{-\al}$ as $\tto$ for $K_j,\al>0$ as in \eqref{pareto}
with  intensity measure $\mu$ as in \eqref{basrakmu} (with $D=\{x\in \overline{\R}^{d}_+:\ \|x\|>1 \}$)  and exponent measure $\nu$ as in \eqref{defMVR} (with $b_n=n^{1/\al}$).
Then  the random vector $F=AV$ as in \eqref{F}  belongs to $\calr(-\al)$.
Moreover, the following assertions hold:\\
(a) \, For all $u>0$
 \beam\label{limitF}
 \frac{\pr{F  \in t \cdot}}{ \pr{\|F\|>t}} \stv  \ov\mu(\cdot),\quad\tto,
 \eeam
 where the intensity measure of $F$ is given by
\beam\label{limitmeasureF}
\ov\mu(\cdot)  = \big(\E \mu \circ A^{-1}(C_{1,\mathbb{S}^{q-1}_{+}})\big)^{-1}\mathbb{E}\mu \circ A^{-1}(\cdot)
\eeam
 on $\calb( \overline \R_+^q\setminus\{\bnull\})$; i.e. $\ov\mu$ is a probability measure on $C_{1,\mathbb{S}^{q-1}_{+}}$. Moreover, $\ov\mu$ is homogeneous of order $-\al$.\\
(b) \,  For all $u>0$
 \beao
 \frac{\P(\|F\|>ut, {F}/{\|F\|}\in \cdot)}{ \P(\|F\|>t)} \stv u^{-\al} \ov\rho(\cdot),\quad\tto,
 \eeao
 where the spectral measure of $F$ is given by
 \beam\label{spectralF}
 \ov\rho(\cdot) = \Big(\E \mu \circ A^{-1}(C_{1,\mathbb{S}^{q-1}_{+}})\Big)^{-1}\mathbb{E}\mu \circ A^{-1}\Big(\Big\{\|x\|>1, \frac{x}{\|x\|} \in \cdot\Big\}\Big),\quad\tto,
 \eeam
 on $\calb(\mathbb{S}^{q-1}_+)$; i.e., it is a probability measure on the sphere $\mathbb{S}^{q-1}_{+}$.\\
\end{proposition}

\proof{Proof:}
(a) \, 
Proposition~A.1 of \cite{Basrak200295} ensures that $F\in\calr(-\al)$ and that
$$\frac{\pr{F\in t\cdot}}{\pr{\|V\|>t}} \stv \E \mu\circ A^{-1}( \cdot),\quad\tto.$$
This implies 
\beao
 \frac{\pr{F \in tu\,\cdot}}{\P(\|F\|>t)} \frac{\pr{\|F\|>t}}{ \pr{\|V\|>t}} \stv 
u^{-\al} \mathbb{E}\mu \circ A^{-1}(\cdot),\quad\tto,
\eeao
as well as 
\beam\label{FV}
\frac{\pr{\|F\|>t}}{\pr{\|V\|>t }} \rightarrow  \E \mu\circ A^{-1}( C_{1,\mathbb{S}_{+}^{q-1}}),\quad\tto.
 \eeam
which gives \eqref{limitmeasureF}. 
Moreover, homogeneity holds, since
\beao
 \E \mu (\{ x: Ax\in u \cdot  \})=\E \mu (u\{ x: Ax\in  \cdot  \})
	&=& u^{-\alpha}\E \mu (\{ x: Ax\in \cdot  \}).
	\eeao	
Part (b) is a consequence of \eqref{limitF} and \eqref{limitmeasureF}.

 \endproof

In what follows we focus on the two scenarios of asymptotically independent  and asymptotically fully dependent claims $V_1,\ldots,V_d$. 
They will play an important role as asymptotic upper and lower bounds for the individual and market risk as we will make precise in Section~\ref{bounds}.
We can now present the tails of the exposure of the individual agents.

\bthe\label{singleFasym}
\begin{enumerate} 
\item  
 For the marginals $F_i=A_{i \cdot}V$, $i=1,\ldots,q$, of $F$ we obtain 
\beam\label{marginscales_ind}
\pr{F_i>t} &\sim&  C t^{-\al}  ,\quad t \to \infty, 
\eeam
 with $C=C^i_{ind}= \sum_{j=1}^{d} K_j\E A_{ij}^{\alpha}$ (as in \eqref{VaRconst}), if $V_1,\dots, V_d$ are asymptotically independent,  and\\
 with 
$C^i_{dep}= \E (AK^{1/\al}\bone)_{i}^{\alpha}$ (as in \eqref{VaRconst_dep}), if $V_1,\dots, V_d$ are asymptotically fully dependent.
\item 
For any subset $\{i_{1},\dots, i_k    \}$ of $\{1,\dots,q\}$  and any choice of positive  $u_{1},\dots, u_k$   we obtain
\begin{gather}\label{jointmargins_ind}
\pr{F_{i_1}>u_1 t,\dots,F_{i_k}>u_k t}\sim \sum_{j=1}^{d} K_j \E \Big(\min \Big\{ \frac{A_{i_{1}j}}{u_1},\dots,  \frac{A_{i_{k}j}}{u_k} \Big\}\Big)^{\alpha}  t^{-\al},\quad t \to \infty,
\end{gather}
in the case of asymptotically independent $V_1,\dots,V_d$, 
and
\begin{gather}\label{jointmargins_dep}
\pr{F_{i_1}> u_1 t,\dots,F_{i_k}> u_k t}\sim    \E \Big(\min \Big\{\frac{ (AK^{1/\al}\bone)_{i_{1}}}{u_1},\dots,  \frac{(AK^{1/\al}\bone)_{i_{k}}}{u_k} \Big\}\Big)^{\alpha} t^{-\al},\quad t \to \infty,
\end{gather} 
in the case of asymptotically fully dependent $V_1,\dots, V_d$.
\item  \label{normFdistr} 
For the aggregated exposures  $\|F\|$ we obtain 
\beam \label{agrregatedtail}
\P(\|F\|> t)\sim   C  t^{-\alpha}, \quad\tto,
\eeam
with $C=C^S_{ind}=\sum_{j=1}^d K_j \E \| A e_j \|^\alpha$  (as in \eqref{VaRconst}), if $V_1,\dots, V_d$ are asymptotically independent,  and\\
  with  $C=C^S_{dep}=\E \|AK^{1/\al}\bone\|^{\alpha}$ (as in \eqref{VaRconst_dep}), if $V_{1},\dots,V_d$ are fully dependent claims.   
\end{enumerate}
\ethe

\proof{Proof:}
The two relations summarized in \eqref{marginscales_ind} are particular cases of \eqref{jointmargins_ind}  and \eqref{jointmargins_dep}.
For a proof of \eqref{jointmargins_ind}  we note first that by \eqref{limitF} and \eqref{FV}
$$
\pr{F_{i_1}>u_1t,\dots, F_{i_k}>u_kt} \sim  \E\mu\circ{A^{-1}} (\{x:\ x_{i_1}>u_1,\dots, x_{i_k}>u_k\})\pr{\|V\|>t}
$$
and proceed, using $\mu=(\sum_{j=1}^d K_j)^{-1}\nu$ from Lemma~\ref{lemma:indep}(c), with
\begin{align*}
\lefteqn{\E\mu(\{x:\ (Ax)_{i_1}>u_1,\dots, (Ax)_{i_k}>u_k\})}\\
&= (\sum_{j=1}^{d} K_j)^{-1} \sum_{j=1}^{d} 
\E \nu^*\Big(\Big\{se_{j}: s>\max\Big\{\frac{u_1}{K_j^{1/\al}A_{i_{1}j}},\dots, \frac{u_k}{K_j^{1/\al}A_{i_{k}j}}\Big\}\Big\}\Big)\\
&=   (\sum_{j=1}^{d} K_j)^{-1}\sum_{j=1}^{d} 
\E\Big(\min  \Big\{ \frac{K_j^{1/\al}A_{i_{1}j}}{u_1},\dots,\frac{ K_j^{1/\al}A_{i_{k}j}}{u_k  } \Big\}\Big)^{\alpha}\\
&=  (\sum_{j=1}^{d} K_j)^{-1}\sum_{j=1}^{d} K_j
\E\Big(\min  \Big\{ \frac{A_{i_{1}j}}{u_1},\dots,\frac{A_{i_{k}j}}{u_k  } \Big\}\Big)^{\alpha}.
\end{align*}   
We find from Lemma~\ref{lemma:indep}(b) that $\pr{\|V\|>t}\sim \nu(\{\|x\|>1\})  t^{-\al} \sim \sum_{j=1}^d K_j t^{-\al}$, which completes the proof of \eqref{jointmargins_ind}.\\
Relation \eqref{jointmargins_dep} is shown analogously incorporating that the respective exponent measure for fully dependent variables is concentrated on   $\{ sK^{1/\al}\bone: s>0\}$, in particular, $\pr{\|V\|>t}\sim \|K^{1/\al}\bone\|^\al t^{-\al}$ as $\tto$. \\
Relation \eqref{agrregatedtail}  follows from Proposition~\ref{th1}(a), if we take into account that 
\begin{gather}\label{connectioncircconst}
\E \mu \circ A^{-1}(C_{1,\mathbb{S}^{q-1}_+})
= (\sum_{j=1}^d K_j)^{-1} C^S_{ind}\quad \text{and}\quad 
\E \mu \circ A^{-1}(C_{1,\mathbb{S}^{q-1}_+})= \|K^{1/\al}\bone\|^{-\alpha}C^S_{dep}\end{gather}
for asymptotically independent and asymptotically fully dependent claims $V_1,\dots,V_d$.
 
\endproof

\gdr{These general results on regular variation are independent of the specific form of the matrix $A$ as given in \eqref{eq2.2}.
When we want, however, to compute the dependence measures like the spectral measure explicitly, the specific form of \eqref{eq2.2} becomes relevant.
If $\deg(j)$ denotes the number of agents insuring object $j$, then for the choice of $A$ as in \eqref{eq2.21}; 
$$
 A_{ij} = \frac{\1 (i\sim j)}{\deg(j)}, \quad \mbox{ where } \frac{0}{0}:=0,
$$
the loss is evenly distributed across all agents insuring object $j$.
For such $A$ the relation \eqref{jointmargins_ind} shows that the joint tail behaviour of $F_i$ and $F_k$ is asymptotically independent  if the corresponding different agents $i$ and $k$ do not insure any object jointly.

If $A$ is as in \eqref{eq2.21}, then
 we can calculate the spectral measure explicitly as follows.}

\begin{corollary}\label{thmrho}
Let the assumptions of Proposition~\ref{th1} hold, \gdr{let $A$ be as in \eqref{eq2.21}}, and assume that the claims $V_1,\dots, V_d$ are asymptotically independent. 
Then  the spectral measure $\ov\rho$ of $F$ has support given by
 \beam\label{supp(rho)}
\operatorname{supp}(\ov\rho)\subseteq S
&=& \left\{ \|x\|^{-1} x  :\  x \in \{0,1\}^q \setminus \{0\} \right \};
\eeam
i.e., $\ov\rho$ is  concentrated on at most $2^q-1$ many points on the sphere.
Setting  $C^S_{ind}=  \sum_{j=1}^d K_j  \E\|A e_j\|^\al$ (as in \eqref{VaRconst}), we get for $b=(b_{1},\dots,b_{q}) \in S$
\begin{gather}\label{rhogenau}
\ov\rho(\{b\})=(C^S_{ind})^{-1}\left\|\frac{(\mathds{1}(b_{1}>0),\dots,\mathds{1}(b_{q}>0))^\top}{|b|_{H}} \right\|^{\alpha} \sum_{j=1}^{d}K_j\prod_{i=1}^{q} p_{ij}^{\mathds{1}( b_{i}>0 )}(1-p_{ij})^{\mathds{1}(b_{i}=0)}.
\end{gather} 
\end{corollary}

\proof{Proof:}
Recall that, by independence, the spectral measure of $V$ is concentrated on the intersections of the sphere with the axes.
Hence, the spectral measure $\ov\rho$ of $F$ is concentrated on 
$$ \left\{ \|M K^{1/\al} e_{j} \|^{-1} M K^{1/\al} e_{j}:\ j=1,\dots,d; Me_{j}\neq 0; \ M \in \operatorname{supp}(A) \right\} \subseteq \left\{ \|x\|^{-1} x  :\  x \in \{0,1\}^q \setminus \{\bnull\} \right \}, $$ 
hence, \eqref{supp(rho)} follows. 
To prove \eqref{rhogenau}, we use that $\nu$ and  $\mu$ are concentrated on the axes and conclude  from Lemma~\ref{lemma:indep}(a),
 \beam 
 \E\mu\circ A^{-1}(C_{1,\mathbb{S}_{+}^{q-1}}) &=&  \E \mu(\{x \in \overline{\R}^d_+  :\ Ax \in  C_{1,\mathbb{S}_{+}^{q-1}}\}) \nonumber \\
  &=& (\sum_{j=1}^d K_j)^{-1}  \E \nu(\{x\in\overline{\R}^d_+ : Ax\in C_{1,\mathbb{S}_{+}^{q-1}}\})\nonumber \\
   &=& (\sum_{j=1}^d K_j)^{-1}   \E\nu^*\circ K^{-1/\al} (\{x\in\overline{\R}^d_+  : \|Ax\| >1\})\nonumber \\
 &=& (\sum_{j=1}^d K_j)^{-1}   \E\nu^*(\{x\in\overline{\R}^d_+  : \|AK^{1/\al} x\| >1\})\nonumber \\
    &=& (\sum_{j=1}^d K_j)^{-1}  \sum_{j=1}^d\E \nu^*(\{te_j:  t\ge (\|AK^{1/\al} e_j\|)^{-1}\}) \nonumber \\
&=&   (\sum_{j=1}^d K_j)^{-1}  \sum_{j=1}^dK_j \E \|Ae_j\|^\al
\,  = \,  (\sum_{j=1}^d K_j)^{-1} C^S_{ind} . \label{indeptcase}
 \eeam
 Now note that for $b \in S,$  
$$
\mu(\{ t e_{j}:\  t > \|Ae_{j}\|^{-1}, \| A e_{j}\|^{-1}A e_{j}=b  \})
=\mu(\{ t e_{j}:\  t > \|Ae_{j}\|^{-1} \}) \mathds{1}\{\| A   e_{j}\|^{-1}A e_{j}=b\}
$$
holds. Consequently, by homogeneity,  for $b \in S,$
\begin{align*}
&\E\mu\circ A^{-1}(\{x\in \overline{\R}^{q}_+:\ \|x\|\geq 1,\ x/\|x\|=b \})\\ &= \E \mu(\{x \in \ov\R^{d}_{+}:\  \|Ax  \|>1,\   \|Ax\|^{-1}Ax=b\})\\
&=\E \sum_{j=1}^{d} \mathds{1}\{\| A   e_{j}\|^{-1}A e_{j}=b\} \mu(\{ t e_{j}:\  t > \|Ae_{j}\|^{-1} \})\\
 &= \E \sum_{j=1}^{d} \mathds{1}\{\| A   e_{j}\|^{-1}A e_{j}=b\}  K_j\|A e_j\|^\alpha (\sum_{j=1}^d K_j)^{-1}.
\end{align*}
Next, observe that under the condition $\| A   e_{j}\|^{-1}A e_{j}=b$ some component of $b$ is not zero if and only if the corresponding component of $A{e_{j}}$ is not zero. This implies
\begin{align*}
&\E\mu\circ A^{-1}(\{x\in \overline{\R}^{q}_+:\ \|x\|\geq 1,\ x/\|x\|=b \})\\ &= \frac{1}{\sum_{j=1}^d K_j} \left\|\frac{(\mathds{1}(b_{1}>0),\dots,\mathds{1}(b_{q}>0))^\top}{|b|_{H}} \right\|^{\alpha} \sum_{j=1}^{d}K_j\pr{\| A   e_{j}\|^{-1}A e_{j}=b}\\
&= \frac{1}{\sum_{j=1}^d K_j}\left\|\frac{(\mathds{1}(b_{1}>0),\dots,\mathds{1}(b_{q}>0))^\top}{|b|_{H}} \right\|^{\alpha}\sum_{j=1}^{d} K_j\prod_{i=1}^{q} p_{ij}^{\mathds{1}( b_{i}>0 )}(1-p_{ij})^{\mathds{1}(b_{i}=0)},
\end{align*}	
and in view of Proposition~\ref{th1}(b) we get the result. 

\endproof

 The support of the spectral measure will also be finite if there is full asymptotic dependence between the claims. 
 
\begin{corollary}
Let the assumptions of Proposition~\ref{th1} hold, \gdr{let $A$ be as in \eqref{eq2.21}}, and assume that the claims $V_1,\dots, V_d$ are asymptotically fully dependent.
Then  the spectral measure $\ov\rho$ of $F$    has support
 \beam\label{supp(rho)_dep}
\operatorname{supp}(\ov\rho) &=& \left\{ \|M K^{1/\al}\bone \|^{-1} MK^{1/\al}\bone:\  M \bone  \neq 0; \ M \in \operatorname{supp}(A) \right\}.
\eeam
Setting $C^S_{dep}=\E\|AK^{1/\al}\bone\|^\al$ (as in \eqref{VaRconst_dep}), it takes on $\calb(\S^{q-1}_+)$ the form
$$\ov\rho(\cdot)=  \|K^{1/\al} \bone\|^{\alpha} (C^S_{dep})^{-1}\mathbb{E}\mu \circ A^{-1}\Big(\Big\{\|x\|>1, \frac{x}{\|x\|} \in \cdot\Big\}\Big). $$  
\end{corollary}

\proof{Proof:}
Since the spectral measure of $V$ is concentrated on $K^{1/\al}\bone$, the spectral measure $\ov\rho$ of $F$ is concentrated on $M K^{1/\al}\bone$ for $j=1,\ldots,d$ and $M\in\supp(A)$, normalized to live on the sphere. This implies \eqref{supp(rho)_dep}. 
 From Lemma~\ref{lemma:indep}(a) with (d)  we conclude,
 \beam\label{constant}
 \E\mu\circ A^{-1}(C_{1,\mathbb{S}_{+}^{q-1}}) &=&  \E \mu(\{x \in\overline\R^d_{+}:\ Ax \in  C_{1,\mathbb{S}_{+}^{q-1}}\})\nonumber\\
& = & \|K^{1/\al}\bone\|^{-\alpha} \E \|A K^{1/\al}\bone\|^\al \, = \, \|K^{1/\al}\bone\|^{-\alpha} C_{dep}^S.
 \eeam
Using this, we find from Proposition~\ref{th1}(d) and \eqref{RV1},
	\begin{gather*}
 \frac{\P(\|F\|>ut, {F}/{\|F\|}\in \cdot)}{ \P(\|F\|>t)} \stv u^{-\al} 
  \|K^{1/\al}\bone\|^{\alpha} (C_{dep}^S )^{-1}
 \mathbb{E}\mu \circ A^{-1}\Big(\Big\{\|x>1\|,\frac{x}{\|x\|} \in \cdot\Big\}\Big),\quad\tto,
	\end{gather*} 
	on $ \mathcal{B}(\mathbb{S}_{+}^{q-1})$, giving the spectral measure $\ov\rho$ as above. 
	
\endproof

\gdr{We end this subsection with a result concerning the tail behaviour of uninsured losses. }

\Ored{
\begin{proposition}\label{prop:uninsured}
Let the assumptions of Proposition~\ref{th1} hold and assume the market matrix entries of the form $A_{ij}= \1(i\sim j)W_{ij}$ as in \OK{\eqref{eq2.21}} with $ 0< \sum_{i=1}^q W_{ij}\leq 1$ for $j=1,\dots,d$. Then the tail behaviour of uninsured losses is given by
\begin{gather}\label{asym:uninsured}
\pr{\sum_{j=1}^d   (1-\sum_{i=1}^q W_{ij}1(i\sim j ))   V_j  \ge t  }\sim t^{-\alpha} B
\end{gather}
 with $B=B_{ind}= \sum_{j=1}^{d}K_j \E (1-\sum_{1=1}^q 1(i\sim j)W_{ij}  )^{\alpha}$ if the object claims are asymptotically independent and $B=B_{dep}= \E( \sum_{j=1}^d K_j^{1/\al}(1- \sum_{i=1}^q W_{ij} 1(i\sim j) )  )^{\alpha} $ if the object claims are fully dependent.
In the special case, where  $A_{ij}= \deg(j)^{-1} \1(i\sim j)$, we have 
$B=B_{\ind}:=\sum_{j=1}^{d}K_j\prod_{i=1}^{q}(1-p_{ij})  $ and
$B=B_{\dep}:= \sum_{(n_1,\dots,n_d)\in \{0,1\}^d} \prod_{j=1}^{d}\prod_{i=1}^q (1-p_{ij})^{n_j}p_{ij}^{1-n_j}(\sum_{j=1}^{d}K_j^{1/\al}n_j)^{\alpha} $
\end{proposition}

\proof{Proof:}
Similar calculations as in the proof of Theorem~\ref{singleFasym} ensure that
the asymptotics \eqref{asym:uninsured} hold with the respective constants $B_{ind}$ and $B_{dep}$. It remains to calculate the expectations for the specific choice $A_{ij}=\deg(j)^{-1}\1(i\sim j)$ in either dependence case which is standard, in particular since we assumed each agent to choose objects independently and independent from the choices of other agents.

\endproof


\brem \label{insured_plus_uninsured}
Whenever $A_{ij}=\deg(j)^{-1}\1(i\sim j)$, then the sytemic constants $C^S_{ind}$ and $C^S_{dep}$ from \eqref{VaRconst} representing the risk of the reinsurance market add with the constants $B_{ind}$ and $B_{dep}$ representing uninsured losses from Proposition~\ref{prop:uninsured}  to the total risk constants $\sum_{j=1}^d K_j$ and $(\sum_{j=1}^d K_j^{1/\al})^{\al}$ from Lemma~\ref{lemma:indep}; i.e.,
\begin{gather*}
C^S_{ind} + B_{ind}=\sum_{j=1}^d K_j\ \ \operatorname{and}\ \ C^S_{dep} + B_{dep}=(\sum_{j=1}^d K_j^{1/\al})^{\al}. 
\end{gather*}
\erem

}
\subsection{Risk management applications}\label{s5}

We now investigate the {individual and market} risk of an insurance market based on the bipartite graph represented by the random matrix $A=(A_{ij})_{i,j=1}^{q,d}$ as in \eqref{eq2.2} with $q$ agents and $d$ objects. 
Recall that an edge between an agent $i$ and an object $j$ exists with probability $p_{ij}.$

The Value-at-Risk (VaR) of a random variable $X$ at confidence level $1-\gamma$ is defined as
\begin{gather*}
\VaR_{1-\gamma}(X):=\inf\{ t \geq 0: \pr{ X>t } \leq \gamma  \},\quad \gamma \in (0,1),
\end{gather*} 
 and the  Conditional Tail Expectation (CoTE) at confidence level $1-\gamma$, based on the corresponding \VaR,  as
\begin{gather*}
\CoTE_{1-\gamma}(X):=\E[X \mid X> \VaR_{1-\gamma}(X)],\quad \gamma \in (0,1).
\end{gather*}

Note that the Conditional Tail Expectation is also called Expected Shortfall. 
We consider risk measures of $F=AV$, where the random matrix $A$ models the network structure of the market. The claim vector $V$ has Pareto-tailed components, which are assumed to be either asymptotically independent or asymptotically fully dependent.
Hence, $F$ is the multivariate regularly varying vector of the joint exposures of the agents in the market.
\OK{In order to assess the market risk we consider VaR and Conditional Tail Expectation of some norm of the exposure vector $F$, where we draw from the axiomatic {set-up} for systemic risk measures of \cite{axiomsystemic}.}
\CK{In a follow-up paper \cite{KKR2} we address conditional risk measures related to CoVaR (cf. \cite{CoVar}) and Marginal or Systemic Expected Shortfall (cf. \cite{Brownlees}). } 

\gdr{Natural choices for norms are  the $r$-norms $\| x \|_r=(\sum_{i=1}^{q} |x_{i}|^{r})^{1/r}$  for $r\ge 1$.
However,  for $r>1$ such a norm can have fatal economic consequences.
Assume that a claim of size $V_j$ is distributed equally between those agents, which insure this particular claim, which corresponds to \eqref{eq2.21}.
Then 
$$\|A_{ij} V_j\|_r =\Big (\sum_{i=1}^q V_j^r \frac{\1(i\sim j)}{\deg(j)^r}\Big)^{1/r}  
= V {_j} (\deg(j))^{1/r-1} {\le} V_j$$ 
with equality for $r=1$ and strict inequality for $r>1$ when $\deg(j) >0$.
This behaviour would allow for regulatory arbitrage - any loss of size $V_j$ could be made smaller by splitting it between more agents. We will discuss this problem further in Section~\ref{other}.}

Instead of attributing a risk measure to an agent's exposure or to the market exposure, we write for short an agent's risk or the market risk. For the \VaR\ we obtain the following.
 
\begin{corollary}\label{Cor:VaR} 
Let $\al>0$ and $F=(F_{1},\dots,F_{q})^{\top}$ the vector of the agents' exposures.\\
The individual Value--at--Risk of agent $i\in\{1,\ldots,q\}$ shows the asymptotic behaviour 
 \begin{gather}\label{uniVaRasym}
\VaR_{1-\gamma}(F_{i})\sim C^{1/\alpha} \gamma^{-1/\alpha} ,\quad \gamma\rightarrow 0,
\end{gather} 
with either $C= C^i_{ind}$ or $C=C^i_{dep}$ in case  $V_1,\dots,V_d$ are asymptotically independent or asymptotically fully dependent.
The market Value--at--Risk of the aggregated vector $\|F\|$ satisfies
\begin{gather}\label{marketVaRasym}
\VaR_{1-\gamma}(\|F\|) \sim C^{1/ \alpha}\gamma^{-1/\alpha},\quad \gamma\rightarrow 0,
\end{gather}
with either $C= C^S_{ind}$ or $C=C^S_{dep}$ in case  $V_1,\dots,V_d$ are asymptotically independent or asymptotically fully dependent.
\end{corollary}

\proof{Proof:}
In view of Theorem~\ref{singleFasym}, the asymptotic results \eqref{uniVaRasym} and \eqref{marketVaRasym} hold, since inverses of regularly varying functions are again regularly varying (cf. Prop.~1.5.15 of \cite{BGT1987}). 

\endproof

The analogous result for the Conditional Tail Expectation reads as follows.

\begin{corollary}\label{Prop:ES}
Let $\alpha >1$ and $F=(F_{1},\dots,F_{q})^{\top}$ the vector of the agents' exposures.   \\
The individual Conditional Tail Expectation of  agent $i\in\{1,\ldots,n\}$ shows the asymptotic behaviour 
$$\CoTE_{1-\gamma}(F_{i}) \sim \frac{\alpha}{\alpha-1} \VaR_{1-\gamma}(F_{i})\sim \frac{\alpha}{\alpha-1} C^{1/\alpha} \gamma^{-1/\alpha}   \,,\quad \gamma\rightarrow 0,$$ 
with either $C= C^S_{ind}$ or $C=C^S_{dep}$ in case  $V_1,\dots,V_d$ are asymptotically independent or asymptotically fully dependent.
The market Conditional Tail Expectation of  the aggregated vector $\|F\|$ satisfies
$$\CoTE_{1-\gamma}(\|F\|) \sim \frac{\alpha}{\alpha-1} \VaR_{1-\gamma}(\|F\|)\sim \frac{\alpha}{\alpha-1} C^{1/ \alpha}\gamma^{-1/\alpha}  \,,\quad \gamma\rightarrow 0,$$ 
with either $C= C^S_{ind}$ or $C=C^S_{dep}$ in case $V_1,\dots,V_d$ are asymptotically independent or asymptotically fully dependent.
\end{corollary}

\proof{Proof:}
Both assertions are consequences of Karamata's Theorem (cf. Theorem~1.6.5 of \cite{BGT1987}) and Corollary~\ref{Cor:VaR}. 

\endproof

\subsection{Bounds for individual and market risk}\label{bounds}
Let $V_{ind}, V, V_{dep}$ be claim vectors, all three having the same margins as in \eqref{pareto}, but $V$ having arbitrary dependence structure ($V_{ind}, V_{dep}$ are as before).
Define constants 
\begin{gather}\label{genconstants}
C_{V}^i= \E\nu\circ A^{-1}(\{ x: x_i>t\})\,, i=1,\ldots,1, \quad\text{and} \quad 
C_{V}^S= \E\nu\circ A^{-1}(\{ x: \|x\|>t\}). 
\end{gather}

The following bounds have been established in \cite{KK}. 
 For the  constants $C^{i}$ referring to agent $i$:
\beao
C^{i}_{ind} \leq  C_V^{i} \leq C^i_{dep} \quad \text{for $\alpha\geq 1$},\label{Calphagreateroneindiv}\\
C^{i}_{dep} \leq C_V^{i} \leq C^i_{ind} \quad \text{for $\alpha\leq 1$}.\label{Calphasmalleroneindiv}
\eeao
For the {{system}} constants $C^S$ the interaction of the $r$-norm with the parameter $\al$ leads to the bounds:
\beao
 C^S_{ind}  \leq  C^S_V  \leq  C^S_{dep} \quad &\text{for } \alpha \geq r, \label{sysconstupper} \\
C^S_{dep}   \leq  C^S_V  \leq  C^S_{ind} \quad &\text{for }  0<\alpha<  1.\label{sysconstupperind}
\eeao
We want to emphasize that for $\al \in (1,r)$ the constants $C^{S}_{ind}$ and $C^S_{dep}$ are in general neither upper nor lower bounds \CK{for  $C^S_V$}; for examples see \cite{KK}.

{These bounds justify concentrating on the independent case and on the fully dependent case as the two extreme cases.} 

\section{Network effects}\label{s53}

In this section we discuss  the influence of the graph structure on the risk, which either individual agents or the system as a whole are exposed to.  For the sake of clarity and since it is  most relevant for the type of insurance market under consideration, we mainly concentrate on the situation that the claim variables $V_{1},\dots,V_d$ are asymptotically independent, but we point out some differences to the fully dependent case in Figure~\ref{indep_vs_dep}. 
Both cases are important as they give bounds for possible individual and market risk. \gdr{Moreover, we use as natural aggregation function  the 1-norm $\|\cdot\| = \|\cdot\|_1$ and $A$ as in \eqref{eq2.21}  so that 
$$A_{ij} = \frac{\1(i \sim j)}{\deg(j)} ; \gdr{\quad  \mbox{ where}\ \frac{0}{0} :=0 }$$ except in Section \ref{other}, where we consider alternative norms and matrices.
}

In view of  Corollaries~\ref{Cor:VaR} and~\ref{Prop:ES}, for a given tail index $\alpha$, insight about the effect of the network structure on  the Value-at-Risk and the Conditional Tail Expectation, the individual as well as the market risk,  is solely contained in the constants  given in \eqref{VaRconst} and \eqref{VaRconst_dep}, \GR{which reduce in the above setting as follows. For the individual risk constants we obtain
\beam\label{1constants_ind}
  C_{ind}^i & =& \sum_{j=1}^d K_j \E  \,\frac{ \1 ( i \sim j) }{\deg(j)^\alpha}
\quad \operatorname{and}\quad  C_{dep}^i = \E \Big( \sum_{j=1}^d K_j^{1/\al}  \frac{ \1 ( i \sim j) }{\deg(j)} \Big)^\alpha, 
\eeam
whereas the systemic constants are computed as
\beam
\label{1constants_S_ind} C_{ind}^S &=& \sum_{j=1}^d K_j \E  \Big( \sum_{i=1}^q  \frac{ \1 ( i \sim j) }{\deg(j)} \Big)^{\alpha} \1( \deg(j) > 0) =\sum_{j=1}^d K_j \P (\deg(j) > 0), 
\\ 
\label{1constants_S_dep} C_{dep}^S & =& \E \Big(  \sum_{j=1}^d K_j^{1/\al} \sum_{i=1}^q \frac{ \1 ( i \sim j) }{\deg(j)}  \Big)^\alpha  = \E  \Big(  \sum_{j=1}^d K_j^{1/\al} \1 (\deg(j) > 0)   \Big)^\alpha 
\eeam}
\CK{Before we discuss some examples, we want 
to present approximations for a large system. }

\subsection{Poisson approximations for the constants and the number of non-insured objects} 

If $q$ and $d$ are large then the expressions $\E A_{ij}^\al$ and $\E \| Ae_{j} \|^\alpha$ appearing in \eqref{VaRconst} may not be easy to evaluate. {Both expressions are expectations of a nonlinear function of the edge indicator variables. } The Poisson approximation below involves possibly  fractional moments of the Poisson distribution which may again not be easy to calculate, but {they can be approximated efficiently by simulation, or by numerical approximation of the convergent series. }

\GR{The next proposition considers the case that all edges are independent in the underlying graph, and $\P( i \sim j) = p_{ij}$.}

\begin{proposition}\label{poissonprop}
Let $A_{ij}= \frac{\1 ( i \sim j)}{\deg(j)} $ be as in \OK{\eqref{eq2.21}}, 
where $\{ \1 (i\sim j), 1 \le i \le d, 1 \le j \le q\}$ are independent Bernoulli variables with $\E \1 (i\sim j) = p_{ij}$. 
For  $\lambda_{j}^i=\sum_{k=1, \ldots, q; k \ne i} p_{kj}$ let $X^i_{j}\sim {\rm Pois}(\lambda^i_{j})$ be a Poisson-distributed random variable with mean $ \lambda_j^i$;  let $\lambda_{j}=\sum_{k=1}^{q}p_{kj}$ and $ X_j \sim {\rm Pois}(\lambda_j)$. 
Then 
\begin{gather} \label{Poisson1}  \left| \E A_{ij}^{\alpha} - p_{ij}\E  (1 + X_j^i)^{-\alpha}  \right| \leq p_{ij} \min\{ 1, (\lambda_{j}^{i})^{-1} \} \sum_{k=1, \ldots, q; k \ne i} p_{kj}^{2},
\end{gather} 
and, for the \GR{$1$}-norm, \GR{
\begin{gather} \label{Poisson2} 
\Big| C_{ind}^S - \sum_{j=1}^d K_j (1 - e^{-\lambda_j}) \Big| \leq \sum_{j=1}^d K_j  \min\{ 1, (\lambda_{j})^{-1} \} \sum_{k=1}^{q} p_{kj}^{2}.
\end{gather}}
\end{proposition} 

\proof{Proof:} \,  
By the independence of the edges, 
$\E A_{ij}^\alpha = p_{ij}   \E (1 + \sum_{k=1, \ldots, q; k \ne i} \mathds{1}(k \sim j))^{-\alpha} .$ 
 We define 
$h^u: [0, \infty) \rightarrow [0,1]; \quad  h^{u}(x)=(1+x)^{-\alpha}.$ 
Note that $\sup_{x\ge 0} | h^u(x)| \le 1$. 
As $\sum_{k=1, \ldots, q: k \neq i} \mathds{1}(k \sim j)$ is a sum of independent Bernoulli variables, we may  invoke Eq.~(1.23), p.~8, from \cite{Barbour_etal1992} to obtain the first assertion. \GR{Similarly, the second assertion follows from $\deg(j)$ being the sum of independent Bernoulli variables and Eq.~(1.23), p.~8, from \cite{Barbour_etal1992}.}

\endproof

In particular, if  $\sum_{j=1}^d K_j  p_{ij} \min\{ 1, (\lambda_{j}^{i})^{-1} \} \sum_{k=1, \ldots, q; k \ne i} p_{kj}^{2}$ is small, then $C_{ind}^i$ is well approximated by $ \sum_{j=1}^d p_{ij} \E (1 + X_j^{i})^{-\alpha}$, and if $ \sum_{j=1}^d K_j \min\{ 1, (\lambda_{j})^{-1} \} \sum_{k=1}^{q} p_{kj}^{2}$ is small,
then $C_{ind}^S$ is well approximated by 
$ \sum_{j=1}^d K_j \E X_j^{\alpha \left( {1}/{r} - 1 \right) }$. 
\Ored{
For $C^i_{dep}$  and $C^S_{dep}$ the Poisson approximation yields that, with the notation of  Proposition \ref{poissonprop} and using a Lindeberg argument,
$$ \Big| \E (A K^{1/\alpha} 1)_i^\alpha -  \E \Big( \sum_{j=1}^d K_{j}^{1/\alpha} \1( i \sim j) (1 + X^i_j)^{-1} \Big)^\alpha \Big| \le d(\sum_{j=1}^d K_j^{1/\alpha})^\alpha \min\{ 1, (\lambda_{j}^{i})^{-1} \} \sum_{k=1, \ldots, q; k \ne i} p_{kj}^{2}$$ 
}
and 
\GR{as the degrees of the objects are independent due to the bipartite structure }
\Ored{
$$\Big| C_{dep}^S  -  \E  \Big(  \sum_{j=1}^d K_j^{1/\alpha}  \1 (X_j > 0)   \Big)^\alpha \Big| \le d  \Big( \sum_{\ell=1}^d K_\ell^{1/\alpha} \Big)^\alpha \sum_{j=1}^d \min\{ 1, (\lambda_{j})^{-1} \} \sum_{k=1, \ldots, q} p_{kj}^{2}.$$}

\GR{
\brem 
If $p_{ij} = p$ is identical for all $i,j$, and if $K_j = 1$ for all $j$, then 
$\lambda_j^i = (q-1)p, \lambda_j = qp$, and
the right hand bound of \eqref{Poisson1} simplifies to
$ \min\{ 1, ((q-1)p)^{-1} \}  (q-1) p^3,
$ 
while  the right hand bound of \eqref{Poisson2} becomes
 $
 d \min\{ 1, (q p)^{-1} \}  q  p^2  . 
$
For approximating $C^i_{ind}$ 
we sum over all agents, and the bound gets multiplied by $d$. 
Consider these Poisson approximations when $d$ or $q$ become large. 
Since $p_{kj}=p$ is constant, 
if  $qp\le 1 $ then the Poisson approximation is good when $p$ is much smaller than $(qd)^{-1/3}$. If $qp>1$ then the Poisson approximation is good when $p$ is much smaller than $d^{-1/2}$. 
The quantity $qp$ has an immediate interpretation in networks. With ${\rm deg}(j)  =\sum_{i=1}^q \1 (i \sim j) $  denoting the degree of an object, $qp$ is the expected degree of an object.
\erem 
}

\medskip 
{\color{black}Proposition \ref{prop:uninsured} gives the behaviour of uninsured losses in terms of the network edge probabilities. 
We can also use a Poisson approximation for the risk exposure through non-insured objects, again using the independence of the edges. 

\begin{proposition} \label{noninsuredsize} 
Let $A_{ij}$ be as in \OK{\eqref{eq2.21}}, 
$
A_{ij} = \frac{\1 (i\sim j)}{\deg(j)} 
$
where $\{ \1 (i\sim j), 1 \le i \le d, 1 \le j \le q\}$ are independent Bernoulli variables with $\E \1 (i\sim j) = p_{ij}$. 
 For $j=1, \ldots, d$ let $\pi_j = \prod_{i=1}^{q} (1 - p_{ij}) $ be the probability that $\deg(j) =0$.  
Let  $(X_1, \ldots, X_d)$ be a vector of independent Poisson variables, where $X_j$ has mean $\pi_j$.  
Then 
\beam\label{poiappr}
 \sup_{t > 0 } \Big| \P \Big(\sum_{j=1}^d  \1 (\deg(j) =0)  V_j  \ge t \Big) - \P \Big( \sum_{j=1}^{d} X_j  V_j  \ge t \Big) \Big| \le \sum_{j=1}^d \pi_j^2 . 
\eeam
Whenever the claims are asymptotically independent, 
\begin{eqnarray*}
\P \Big( \sum_{j=1}^{d} X_j  V_j  \ge t \Big)  
&\sim & 
    t^{-\alpha} \sum_{j=1}^d  K_{j} \E   X_j^\alpha,
\end{eqnarray*}
and when the object claims are asymptotically fully dependent, 
\begin{eqnarray*}
\P \Big( \sum_{j=1}^{d} X_j  V_j  \ge t \Big)  
&\sim & 
   t^{-\alpha} \E \Big(\sum_{j=1}^d  K_{j}^{1/\al} X_j\Big)^{\al}.
\end{eqnarray*}
\end{proposition}

\proof{Proof:} \,  Due to the bipartite structure of the graph the indicators $ \1 ({\rm deg}(j) =0)$ are independent for $j=1, \ldots, d$. Let ${\cal{W}} = ( \1 (\deg(j) =0, j=1, \ldots, d) $ be the vector of indicator variables for the non-insured objects. Then from Theorem~10.A , p.~210 in~\cite{Barbour_etal1992}, 
$$ \sup_{0 \le f \le 1} | \E f( {\cal{W}}) - \E f (X_1, \ldots, X_d)| \le  \sum_{j=1}^d \pi_j^2 .$$
Here the functions $f$ are measurable functions with values in the interval $[0,1]$. Conditioning on $V_j, j=1=d$, and using the function 
$f(x_1, \ldots, x_d) = \1 \big( \sum_{j=1}^{d} x_j  V_j  \ge t \big)$ gives the first  assertion. 

Due to the independence between the claim sizes and the network, by the law of total probability, 
\begin{eqnarray*}
\P \Big( \sum_{j=1}^{d} X_j  V_j  \ge t \Big)  
&=& 
 \sum_{n_1, \ldots, n_d =0}^\infty  \Big( \prod_{j=1}^d \P( X_j  = n_j) \Big) \P \Big( \sum_{j=1}^{d} n_j  V_j  \ge t  \Big) . 
\end{eqnarray*}
The standard asymptotics for Pareto claims can now be employed, similar as in the proof of Theorem~\ref{singleFasym}.

\endproof

 \brem
 \begin{enumerate}
\item The quantity $\E \big(\sum_{j=1}^d  K_{j}^{1/\al} X_j \big)^{\al}$ can easily be approximated numerically by standard methods. If the claims  $V_1, \ldots, V_d$ are  exchangeable, so that in particular $K_j=K$ are the same for $ j=1, \ldots,d$,
the sum $\sum_{j=1}^d   X_j$ has Poisson distribution with mean $\lambda = \sum_{j=1}^{d} \pi_j$.
 \item Note that $\sum_{j=1}^d \pi_j$ is the expected number of non-insured objects. The Poisson approximation is good if none of the $\pi_j$'s is very large. For example, if $\pi_j \approx d^{-1}$ for all $j=1, \ldots, d$, then the bound in the Poisson approximation \eqref{poiappr} is of order $d^{-1}$. If $p_{ij} = p_j$ so that the edge probabilities depend only on the object to be insured but not on the agent, then $\pi_j = (1 - p_j)^q $ and the bound in the Poisson approximation is 
 $\sum_{j=1}^d (1 - p_j)^{2q}$. 
 In particular  if $p_j \approx 1 - d^{-\frac{1}{q}}$,  then $\pi_j \approx d^{-1}$. 
\item 
From the perspective of a regulator
the size of losses occurring in the uninsured objects should be small.
If the market allows for a too high amount of uninsured losses, it has failed to work properly.
Proposition~\ref{noninsuredsize} gives an indication for the probability of such market failure for a large market. {{Not only does the number of noninsured losses play a role, but the magnitudes of the losses in terms of the $K_j$'s come into play.  }} 
\end{enumerate} 
\erem
}
  For the number of claims, using Eq.~(1.8), p.~4, in~\cite{Barbour_etal1992}, a Poisson approximation for the number  $W = \sum_{j=1}^{d} \1 (\deg(j) =0) $ of non-insured objects is immediate under the assumptions of Proposition~\ref{poissonprop}.  
Let 
$\lambda = \sum_{j=1}^{d} \pi_j$  and $ X$ Poisson distributed with mean $\lambda$. 
Then 
\begin{equation}\label{noninsurednumber} 
\sup_{A \subset \mathbb{Z}} | \P(W \in A) - \P(X\in A) | \le \min\{ 1, \lambda^{-1}\} \sum_{j=1}^d \pi_j^2. 
\end{equation}

For the approximation \eqref{noninsurednumber} of the number of non-insured objects,  
assume that there is a constant $c \le q$ such that $p_{ij} = \frac{c}{d}$ \GR{for all $i,j$}; every object is equally likely to be insured, but the total expected number of insurance links is kept at $c$, constant, while $q$ may increase to infinity. Then the expected number of non-insured objects is, for large $d$,
$
\E W = d \left( 1 - \frac{c}{d}\right)^q \approx d e^{- \frac{cq}{d}}. 
$
This expression tends to \CK{0 with $q$} even if $q$ is of the same order as $d$.

 When the number of agents increases faster than the number of objects, then there are different limiting behaviours possible.
If, for instance, $q = \alpha d \log d$ for some $\al>0$, then the expected number of uninsured objects tends to zero with $d \rightarrow \infty$ if $\alpha > \frac{1}{c}$ and tends to infinity of $\alpha \le \frac{1}{c}$. 
The bound on the Poisson approximation is at most $\left( 1 - \frac{c}{d}\right)^{2q}$ and will tend to 0 in both cases. 
\CK{Moreover, by  Proposition~\ref{prop:uninsured}, the probability of a large amount of uninsured losses scales for $t\to\infty$ as
$
\P\big(\sum_{j=1}^{d}\mathds{1}(\deg(j)=0)V_j>t\big) \sim t^{-\al} d(1-p)^{q}.
$ 
} 

For the remainder of this section we  first study some exemplifying network situations  and then highlight   the observed common features.
 
\subsection{Examples}\label{examples}\label{s532}

Since our interest is in the graph structure, we take $K_j=1$ in all examples, such that the constants  in \OK{\eqref{1constants_ind} as well as \eqref{1constants_S_ind} and \eqref{1constants_S_dep}  reduce further to}
$$C^i_{ind}=\sum_{j=1}^d \E A_{ij}^\al,\ i=1,\ldots,q,\quad\mbox{and}\quad C^S_{ind}=\sum_{j=1}^d \E\|A e_j\|^\al\GR{ =\sum_{j=1}^d \P ({\rm deg}(j) > 0)},$$
if the claim variables $V_{1},\dots,V_d$ are asymptotically independent, and to
$$C^i_{dep} =\E(A\bone)^\al_i,\ i=1,\ldots,q,\quad\mbox{and}\quad C^S_{dep}=\E\|A\bone\|^\al \GR{=\E  \Big(  \sum_{j=1}^d   \1 ({\rm deg}(j) > 0)   \Big)^\alpha }$$ 
for asymptotically fully dependent claim variables. The larger these constants are, the larger are the Value-at-Risk and the Conditional Tail Expectation.

We discuss three quite different models for the probability matrix $P=(p_{ij})_{i,j=1}^{q,d}$ of edge probabilities and investigate the corresponding risk measures specified by the constants in \eqref{VaRconst} and \eqref{VaRconst_dep}. The first model is a toy model which interpolates between complete specification (one agent insures exactly one object) and full diversification (all agents insure all objects). The second model assumes that all edge probabilities are identical, while the third model is a Rasch-type model, assuming that the edge probabilities result as a product from the propensity of an agent to insure objects and the attractivity of an object. Further models are plausible; exogeneous variables such as location may play a role in that some agents may prefer to insure objects which are in the same geographical location. 
Of course, our framework also applies to deterministic graphs; i.e., $p_{ij}\in \{0,1\},$ and the formulae naturally simplify. 

\subsubsection{A toy model}\label{toysection} 

As a motivating and introductory example we regard a market structure with three agents and three objects, where the random network is specified by the probability matrix
\begin{gather}\label{1bbsquare}
P=\begin{pmatrix}
1& b& b^{2} \\ b^{2} & 1& b \\ b& b^{2 } &1 
\end{pmatrix}. 
\end{gather}
for $b\in[0,1]$.
Observe that the market is completely specified if $b=0$ and fully diversified if $b=1.$ 

\GR{In this example every object has degree at least 1 by construction, and hence  with $d=3$, 
$$ C_{ind}^S 
 =3 , \gdr{\quad  \mbox{ and }  \quad C_{dep}^S = 3^\alpha.} 
$$}

The constants $C^i_{ind}$  and $C^i_{dep}$ can be regarded as functions of $b$.
We can determine expectations by considering all possible realisations of the corresponding vectors, and we do so for $C_{ind}^{i}$ leaving out  $C_{dep}^i$ since calculations are just tedious here. 
Because of the structure of the matrix $P$ we know that $C^i_{ind}$ are identical for $i=1,2,3$.
For agent 1 we get
\beam\label{uniVaRconst1bbsquare}
C^{1}_{ind} &=& \sum_{j=1}^{3} \E A_{1j}^{\alpha}
 = (1-b^2)(1-b) + 2\frac1{2^\al} ( (1-b^2) b + b^2(1-b) ) + 3\frac1{3^\al}b^3\nonumber\\
&=& 1 +b(2^{1-\alpha}-1)+b^{2}(2^{1-\alpha}-1)+  b^{3}( 3^{1-\alpha} + 1 - 2^{2-\alpha})
\eeam
Note that \eqref{uniVaRconst1bbsquare} decreases in $b$ if $\alpha>1$, and it increases if $\alpha <1$. 
This means, from an individual agent's perspective, a more diversified network structure is preferrable for $\al>1$. 
Vice versa, if $\alpha<1$ diversification makes things worse. 
This different impact of diversification has already been observed in several related situations; cf. \cite{ELW, Ibragimov2005, MR, RK99, Zhou}.
In contrast,  such a clear distinction cannot be observed for the market Value-at-Risk constant. \GR{Since every object is insured, the systemic constants do not take further connectivity into account. } 

\subsubsection{Homogeneous probabilities}

Now we assume that the probability matrix $P$ has identical entries; i.e., $P=(p)_{i,j=1}^{q,d}$ and we
study the behaviour of $C^i_{ind}$ and $C^S_{ind}$ as functions of $p\in [0,1]$. By homogeneity $C^i_{ind}$ is the same for all $i$.
We compute
$$
C^i_{ind} = \sum_{j=1}^{d} \E A_{ij}^{\alpha}  = d \, p \sum_{l=0}^{q-1} (1 + l)^{-\alpha} \binom{q-1}{l} p^l (1-p)^{q-1-l}       
$$
and  with \Ored{ $s= 1- (1-p)^{q} = \P  ({\rm deg}(j) > 0) $, 
 
$$ C_{ind}^S = d  s \quad \mbox{ and } \quad 
C_{dep}^S = \sum_{\ell=1}^d {d \choose \ell} \ell^{\alpha} s^{\ell } (1-s)^{d-\ell}  .
$$ 
}
\Gr{Moreover, with $Z_j, j=1,\ldots, d$ i.i.d. Binomial$(q-1,p)$ variables which are independent of 
$ \1  (i \sim j) $ we can write 
$$C_{dep}^i = \E \Big( \sum_{j=1}^d \frac{ \1  (i \sim j)}{ \1  (i \sim j) + Z_j}  \Big)^\alpha. $$}

\Gr{In comparison to \GR{the toy example in Subsection \eqref{toysection}} we observe 
a different behaviour of $C^S_{ind}$}. 
 While in the toy example, 
\GR{increasing $b$ had no effect on the systemic constants, for the homogeneous case the constants are  increasing in $p$.} 
This can be explained as follows.
In \GR{the toy example}
every object is insured with probability $1$ and the diversification effect starts from the beginning. 
In contrast, for homogeneous probabilities and small $p$ only few objects may be insured at all.
Increasing $p$ increases the number of insured objects, assigning also more risk to all agents.
This behaviour  is \GR{illustrated in Figure~\ref{indep_vs_dep} (b) for both $C^S_{ind}$ and $C^S_{dep}$}. 

\begin{figure}
\begin{center}
\includegraphics[width=0.51\textwidth]{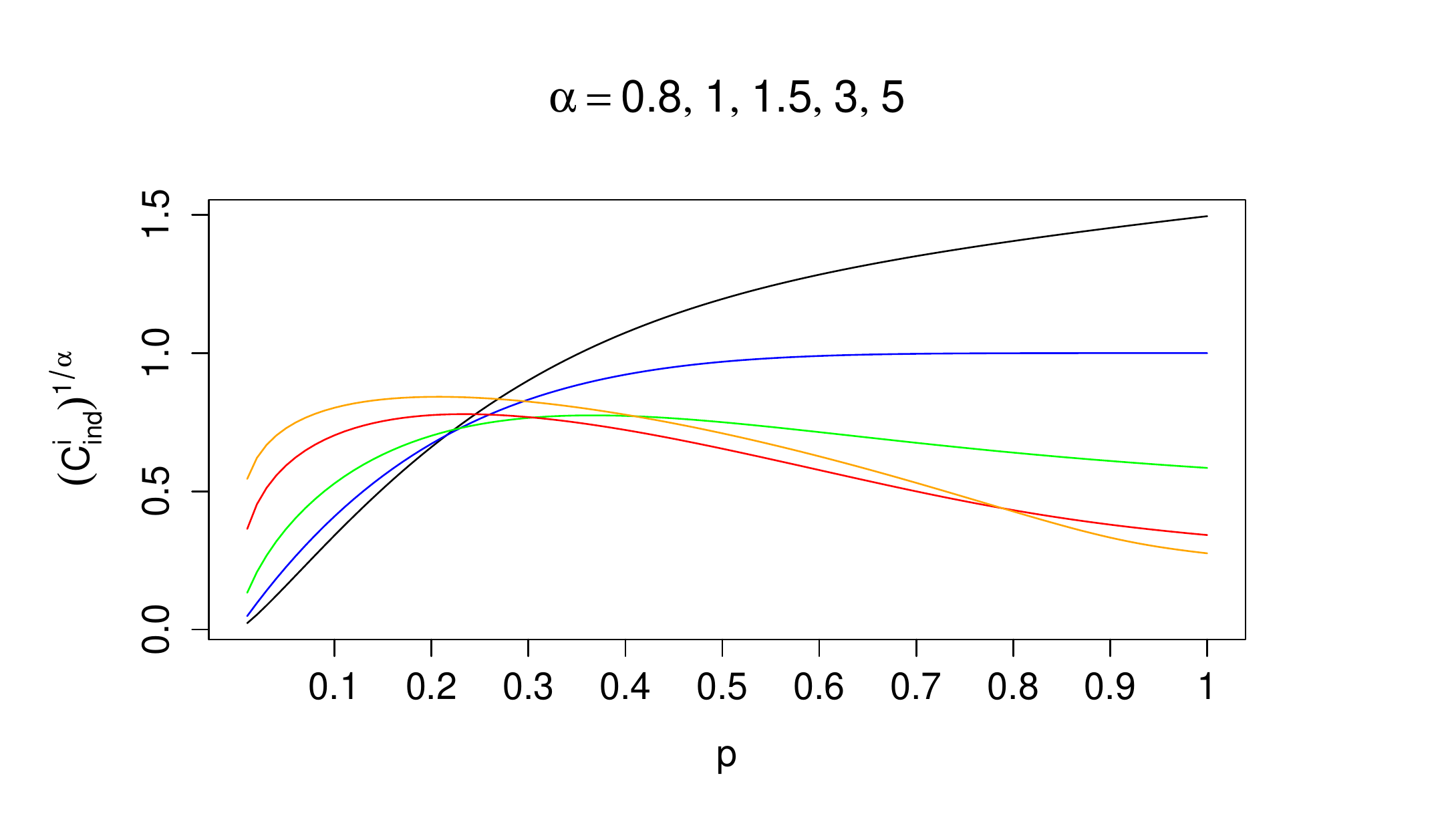}
\vspace*{-0.5cm}
\caption{\label{homonunidiffalphs} Homogeneous probabilities:  $(C^i_{ind})^{1/\alpha}$ with black for  $\alpha=0.8$, blue for $\alpha=1$, green for $\alpha=1.5$, red for $\alpha=3$, orange for $\alpha=5$.}
\end{center}
\end{figure}

\begin{figure}[t]
\subfigure[single agent]{\includegraphics[width=0.49\textwidth, height = 163pt]{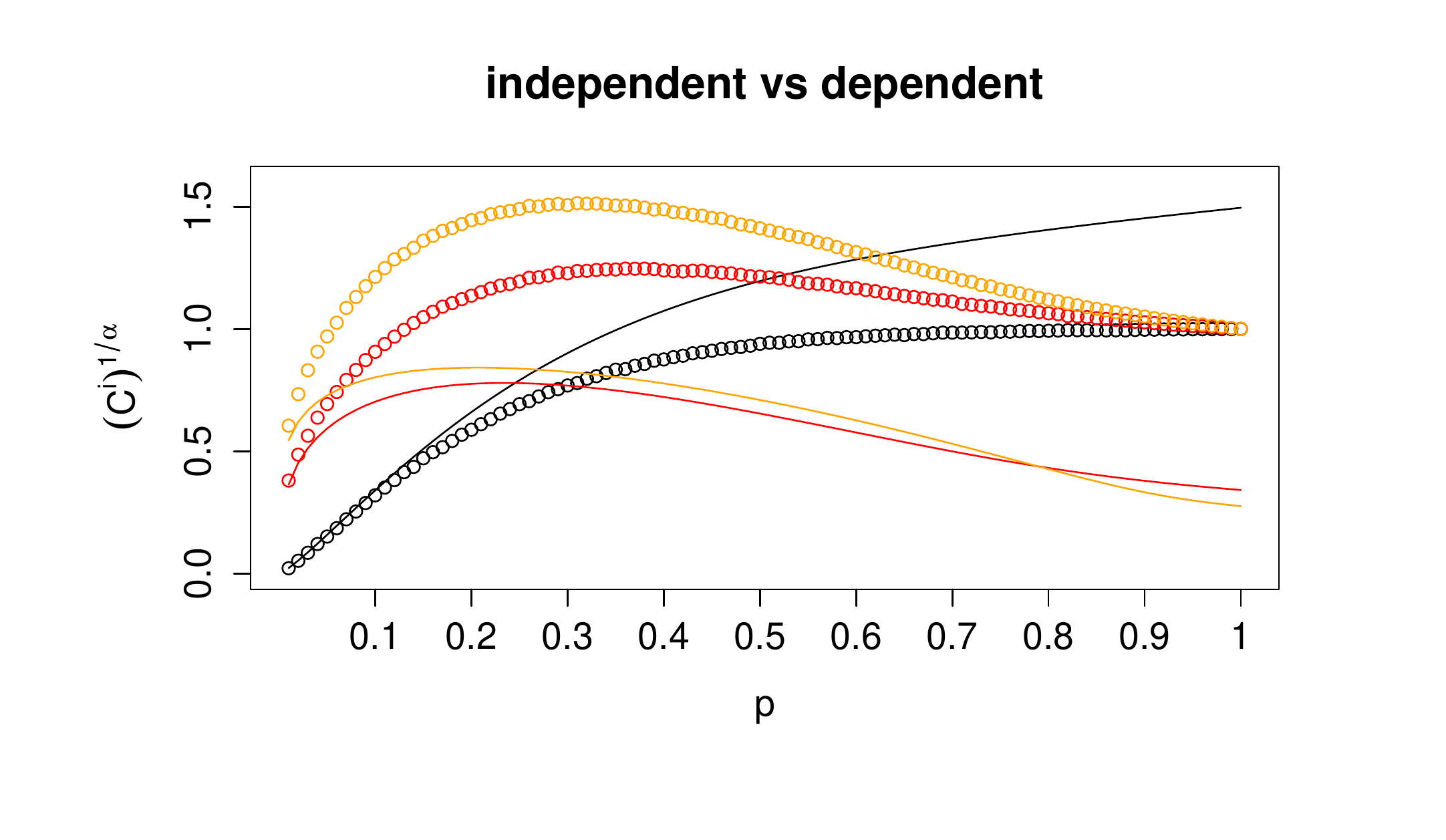}}\hfill
\subfigure[market]{\includegraphics[width=0.49\textwidth]{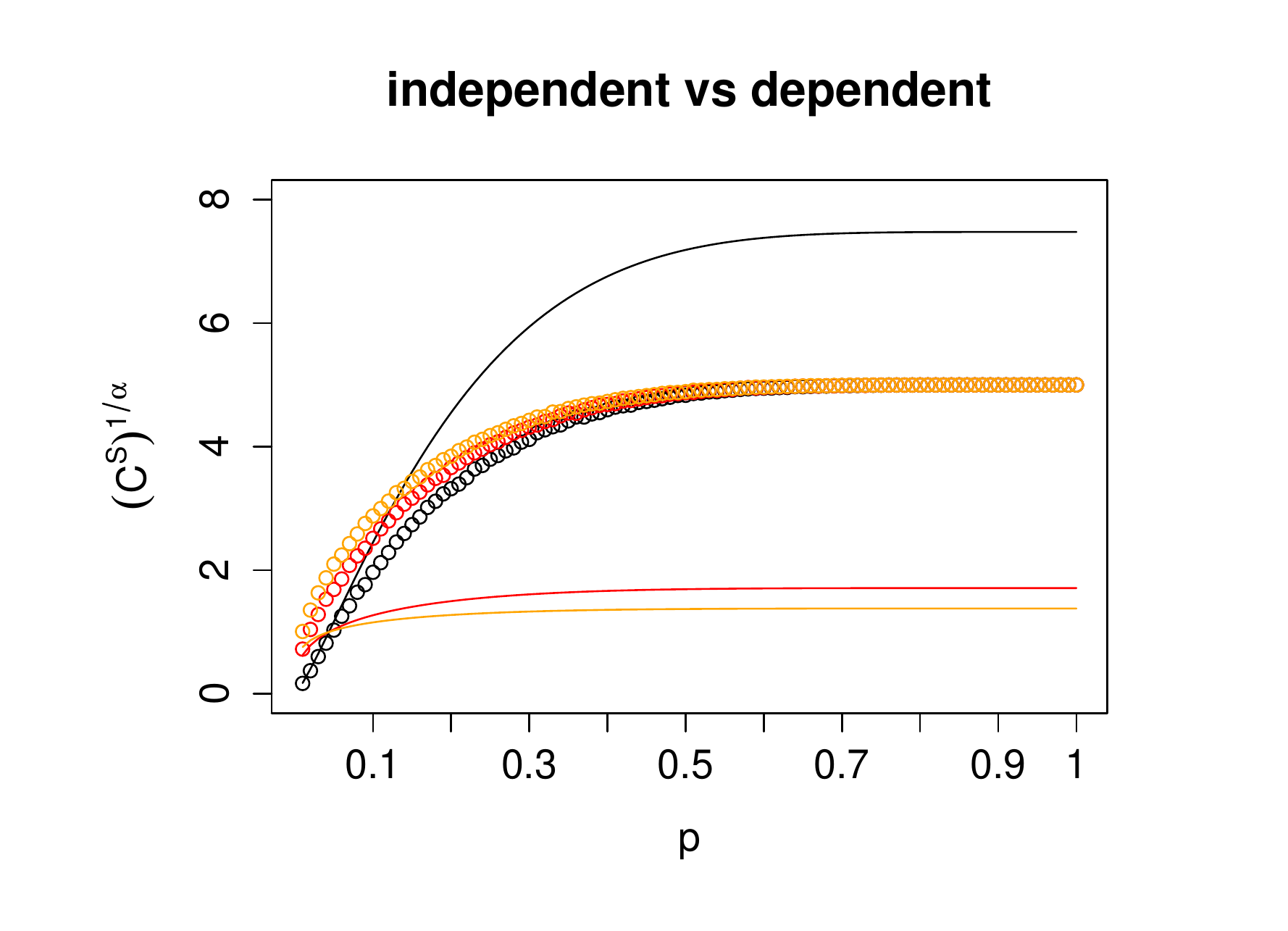}}
\caption{\label{indep_vs_dep} Homogeneous probabilities: Dotted curves depict $(C_{dep}^{i})^{1/\al}$ and $(C^S_{dep})^{1/\al},$ while the solid lines show $(C_{ind}^{i})^{1/\al}$ and $(C^S_{ind})^{1/\al}$ with black for $\alpha=0.8$, red for $\alpha=3$ and orange for $\alpha=5.$ }
\end{figure}

In Figure~\ref{homonunidiffalphs}  we explore the influence of different choices of the tail index $\alpha$ on $C^i_{ind}.$ 	
We recognize that diversification has a beneficial effect for higher values of $p$ only for $\alpha>1.$
In particular, we see that a larger tail index (a lighter tail) causes higher values of $C_{ind}^i$ \gdr{(higher risk)} for  \gdr{very} sparse networks, i.e. for \gdr{very} low values of $p$.  The severity of smaller tail indices $\al$ only arises for higher connected networks. 

At this point, let us again consider the constants $C^i_{dep}$ and $C^S_{dep}$; i.e., the case of fully dependent claim variables. Figure~\ref{indep_vs_dep} compares both types of network constants depending on the parameter $p\in [0,1]$, on the left hand from a individual agent's perspective and on the right hand  from a {{whole system}} perspective. 
A clear difference between the fully dependent and the independent case is that the extent to which the tail index $\alpha$ influences the $\VaR$ constants vanishes as the network connectivity increases. \GR{
We also recognize that over the full range of $p\in [0,1]$, the network constants $C^i_{dep}$ and $C^S_{dep}$ are larger for larger tail indices $\alpha$. This  relationship holds for independent claims only in the range of intermediate connectedness and reverses as connectivity grows.  We can also relate Figure~\ref{indep_vs_dep} to the inequalities in Section~\ref{bounds}. For $\al>1$, independent of the particular market situation reflected by $A$, we have $C^i_{dep} \geq C^i_{ind}$ and 
$C^S_{dep} \geq C^S_{ind}$ whereas for $0<\alpha<1$ the reverse inequalities hold true. 
}

\subsubsection{A Rasch-type model} 

Random bipartite graphs can also be related to the Rasch  models (cf. \cite{rasch})
in social science. These are given by taking
$p_{ij} = \frac{\beta_i\delta_j}{1+\beta_i\delta_j}.$
As a first order approximation of the probabilities in the Rasch model we assume the $p_{ij}$ to be of the form $$p_{ij}=\beta_{i}\delta_{j}$$ for suitable $\beta_{i},\delta_{j} >0$. 
The parameter $\beta_i$ gives a measure for the risk proneness of agent $i$, while the parameter $\delta_{j}$ reflects the attractiveness of object $j$.  

For the random bipartite graph with $p_{ij}=\beta_i\delta_j$ we have
\GR{
\beam \label{raschnorm}
C_{ind}^S =\sum_{j=1}^d \P ({\rm deg}(j) > 0 )  =\sum_{j=1}^d \Big( 1 - \prod_{i=1}^q (1 - \beta_i \delta_j) \Big)
\eeam}
and for $1\leq i \leq q$, 
\beam
\GR{C^i_{ind}}=\sum_{j=1}^{d} \E A_{ij}^{\alpha} &=& \sum_{j=1}^{d} \beta_i\delta_j\sum_{l=0}^{q-1} (l+1)^{-\alpha}\sum_{b \in \{0,1\}^{q}, \atop b_i=1, \|b\|_{1}=l+1} \,\prod_{k=1,\atop k\neq i}^{q}(\beta_k\delta_j)^{\mathds{1}\{ b_{k}>0 \}}(1-\beta_k\delta_j)^{\mathds{1}\{b_{k}=0\}}.\label{raschuni}
\eeam
\CK{Moreover, by  Proposition~\ref{prop:uninsured}, the probability of a large amount of uninsured losses scales as
\begin{gather*}
\P\Big(\sum_{j=1}^{d}\mathds{1}(\deg(j)=0)V_j>t\Big) \sim t^{-\al} \sum_{j=1}^{d} \prod_{i=1}^{q}(1-\beta_i\delta_j).
\end{gather*}
}
We discuss two particular cases.

\begin{example}[Uniform risk proneness of agents]\rm
We take $p_{ij}=\beta_{i}\delta_{j}$  constant in $i$; i.e., $p_{ij}=\beta\,\delta_{j}$.
Then the formulae \eqref{raschnorm} and \eqref{raschuni} simplify to 
\GR{
\beao
C_{ind}^S =\sum_{j=1}^d \Big( 1 - (1 - \beta \delta_j)^q \Big) 
\eeao
and} 
 for $1\leq i \leq q$, 
\beao
\GR{C^i_{ind}=}\sum_{j=1}^{d} \E A_{ij}^{\alpha} &=& \sum_{j=1}^{d}\beta \delta_j\sum_{l=0}^{q-1} (l+1)^{-\alpha} \binom{q-1}{l}(\beta\delta_j)^{l} (1-\beta\delta_j)^{q-1-l}.
\eeao

\begin{figure}
\vspace*{-0.5cm}
\subfigure{\includegraphics[width=0.49\textwidth]{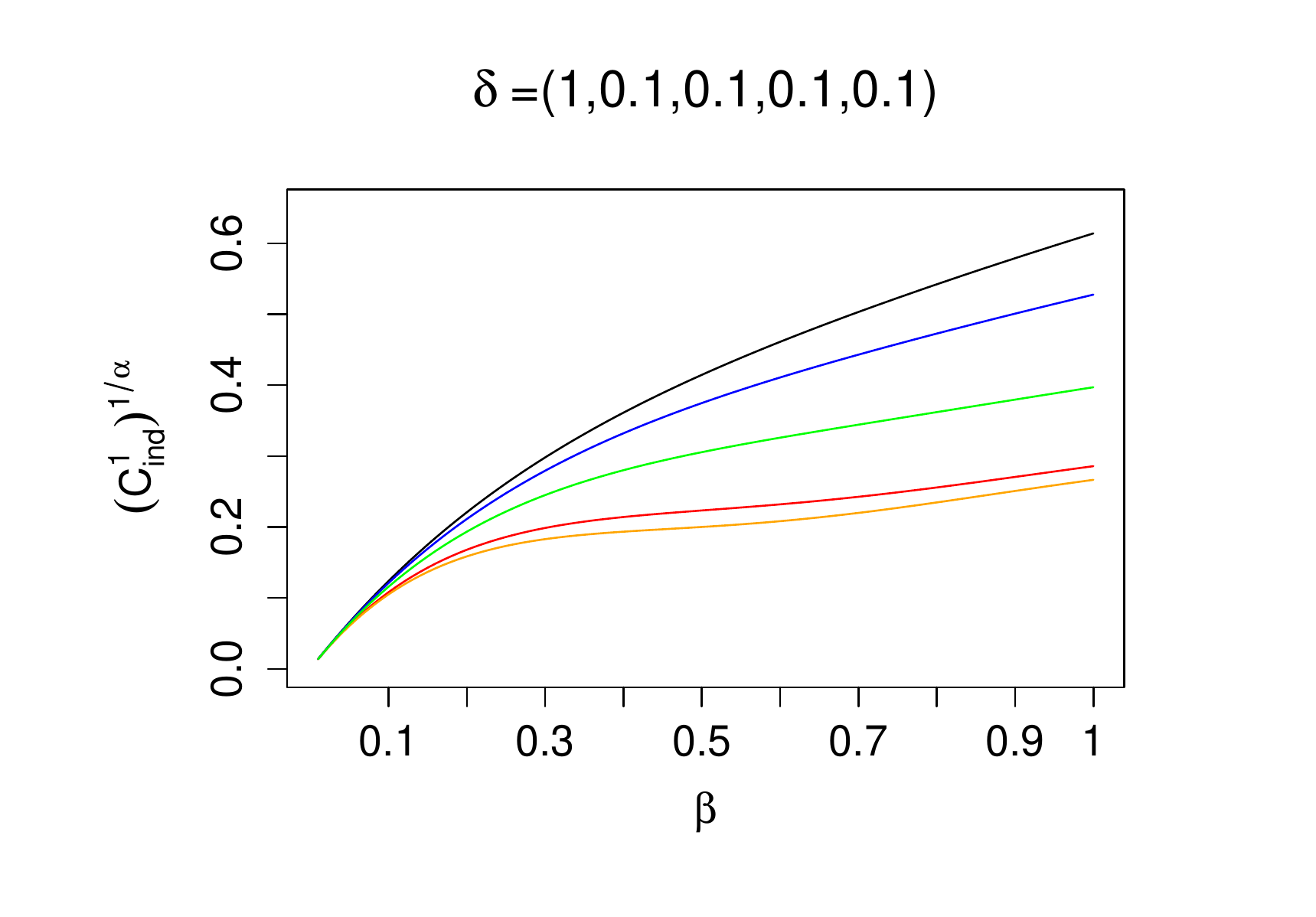}}\hfill
\subfigure{\includegraphics[width=0.49\textwidth]{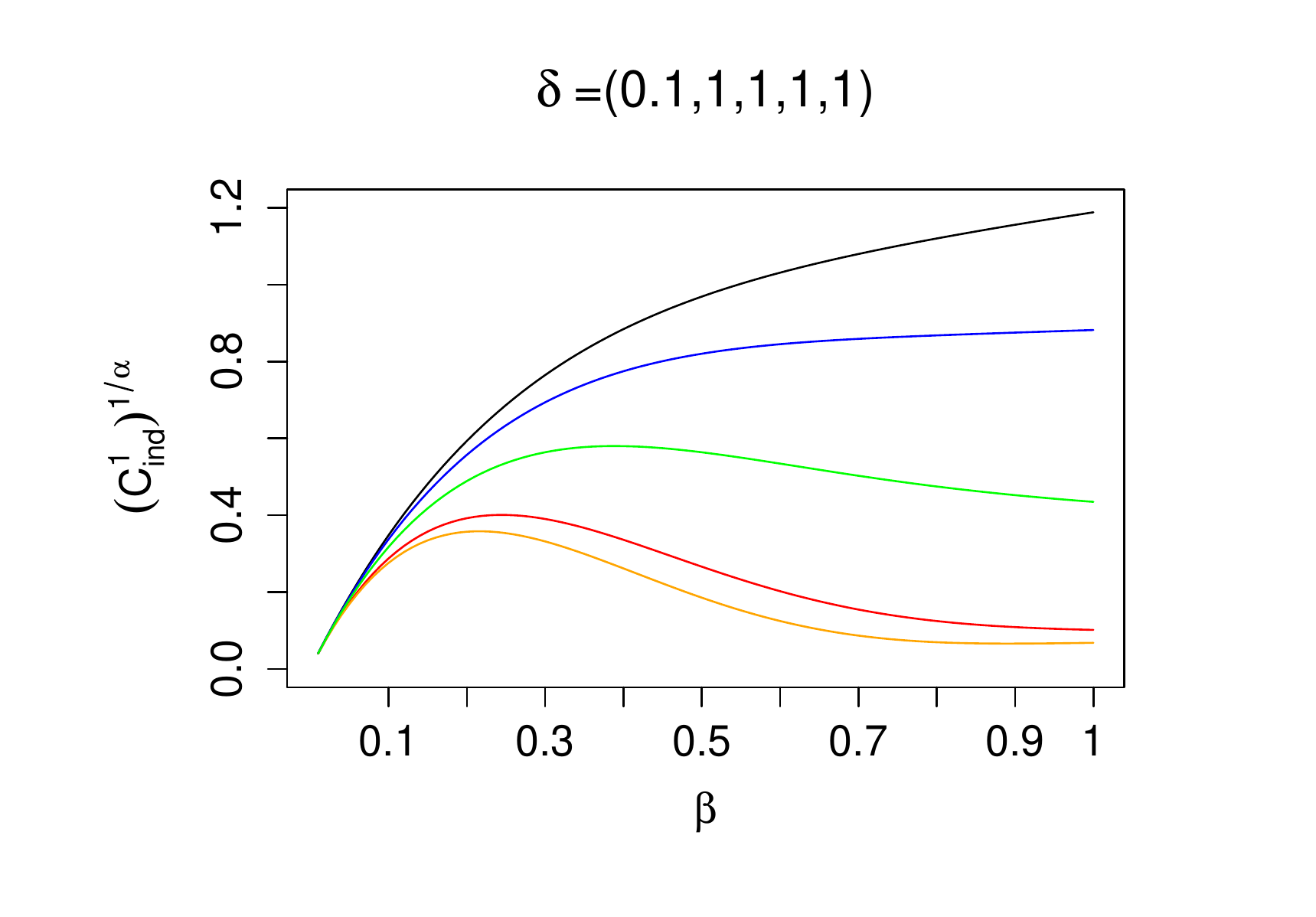}}
\caption{\label{uni-beta-running}  Rasch model:  $(C^1_{ind}(\beta))^{1/\al}$ for different $\alpha$: black for $\al=0.8$, blue for $\al=1$, green for $r=1.5$, red for $r=3$, orange for $\alpha=5$. }
\end{figure}
\GR{
For an illustration see Figure~\ref{uni-beta-running}, where  for the left hand plot we assume a market situation where one object is dominantly attractive in contrast to all the others;  we choose $\delta=(1,0.1,0.1,0.1,0.1)$, while for the right hand plot, we have a market in which only one object is unattractive opposed to the rest which is very attractive; we choose  $\delta=(0.1,1,1,1,1)$. Increasing the connectivity of the network by increasing the  parameter $\beta$ of risk proneness of the agents uniformly leads to no actually observable effect of risk sharing in the left hand plot in the sense of finally decreasing curves. 
This is in contrast to the right hand plot where we can observe beneficial effects of risk sharing for $\al >1$ because a significant number of the objects have the same attractiveness leading to a market situation quite similar to the homogenous case.   
}

\end{example}

\begin{example}[Uniform attractiveness of objects]\rm
We take  $p_{ij}=\beta_{i}\delta$  constant in $j$; i.e., $p_{ij}=\beta_{i}\,\delta$.
Then the formulae \eqref{raschnorm} and \eqref{raschuni} simplify to 
\GR{
\beao
C_{ind}^S  =\sum_{j=1}^d \Big( 1 - \prod_{i=1}^q (1 - \beta_i \delta) \Big)
\eeao}
and for $1\leq i \leq q$, 
\beao
\GR{C^i_{ind}}=\sum_{j=1}^{d} \E A_{ij}^{\alpha} &=&  {\sum_{j=1}^{d}} \beta_i\delta\sum_{l=0}^{q-1} (l+1)^{-\alpha} \, \delta^{l}(1-\delta)^{q-1-l} \sum_{b \in \{0,1\}^{q},\atop b_i=1, \|b\|_{1}=l+1} \prod_{k=1, \atop k\neq i}^{q}(\beta_k)^{\mathds{1}\{ b_{k}>0 \}}\left(\frac{1-\beta_k\delta}{1-\delta}\right)^{\mathds{1}\{b_{k}=0\}}.
\eeao

Figure~\ref{uni-delta-running} gives some insight in the univariate case. If only one key player is present, this key player will not benefit from risk sharing, while the players in the rather inactive group finally experience beneficial effects. 
\end{example}

\begin{figure}
\vspace*{-0.5cm}
\subfigure{\includegraphics[width=0.49\textwidth]{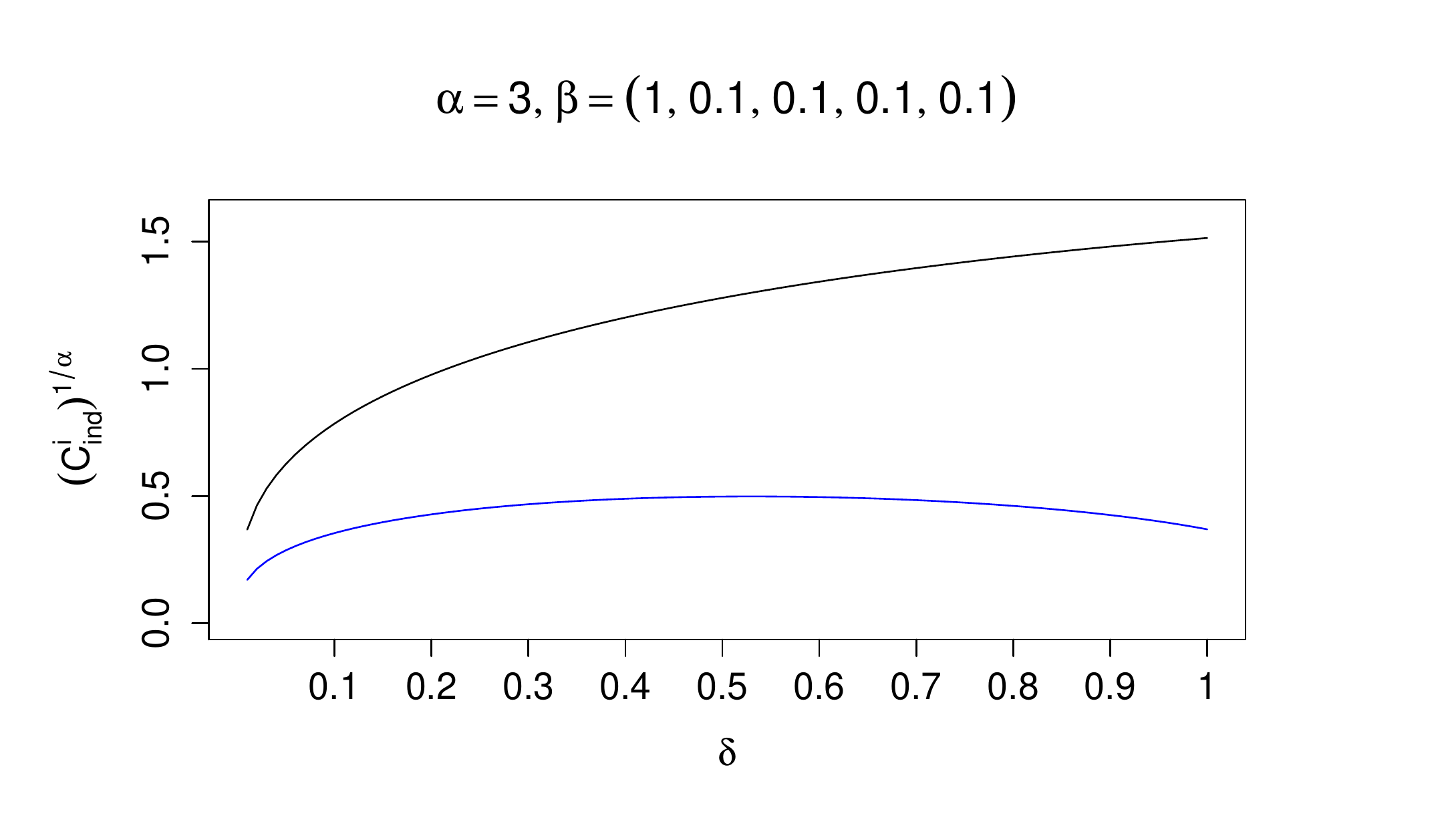}}\hfill
\subfigure{\includegraphics[width=0.49\textwidth]{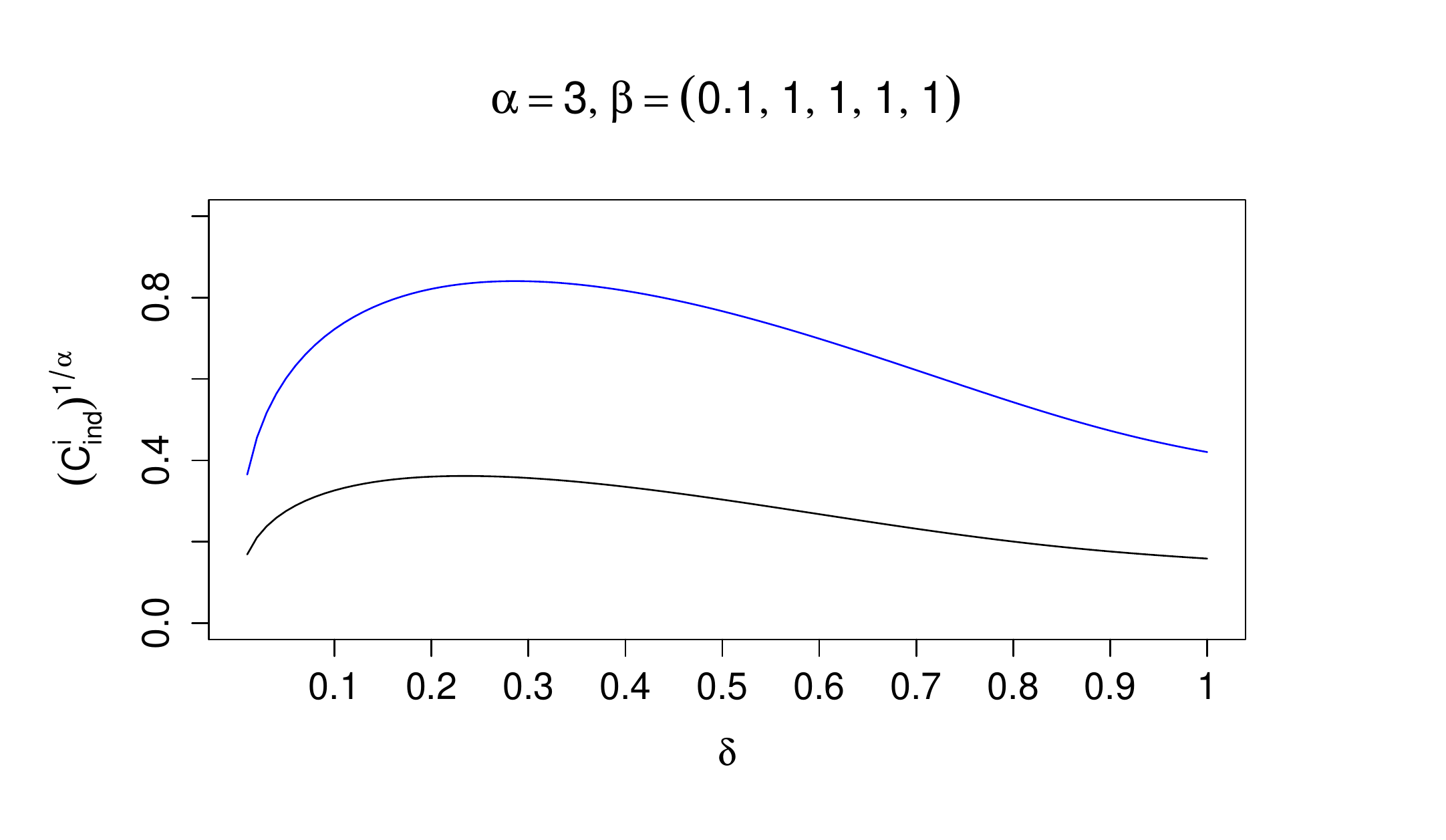}}
\vspace*{-0.4cm}
\caption{\label{uni-delta-running} Rasch model: $(C^i_{ind}(\delta))^{1/\al}$ for different agents: black for $i=1$, blue for $i=2$.}
\end{figure}

\subsubsection{Degree dependence}

One might wonder whether the constants in \eqref{VaRconst} may be described in terms of the summary statistic of average object degree, even in a setting when the network grows. The following example shows that this is not the case. 

We consider two different networks which are deterministic with $d=q$ (i.e. number of agents is equal to number of objects) and given by  their (bipartite) adjacency matrices. The adjacency matrices are called $\wt A_{1}$  and  $\wt A_{2}$, given in Table \ref{matrices}.  For the sake of our argument we  assume that $q $  is divisible by 3.

\begin{table} 
\begin{center}
\begin{tabular}[h]{lc}
 {$\wt{A}_1=\left(
\begin{array}{cccccc}
	1 & 1& 0& \cdots & 0 & 0  \\
	 0 & 1& 1 &  0 & \vdots &\vdots \\
	\vdots & \ddots  & \ddots &\ddots &\ddots & \vdots \\
	0& \cdots& 0 &  1  & 1& 0      \\
	0 & 0& 0& \cdots & 1 & 1 \\
	1 & 0& 0& \cdots & 0 & 1 
\end{array}
\right)$} \quad & \quad \quad \quad \quad \quad \quad \quad  {$ \wt{A}_2=\left(
\begin{array}{ccc}
	\begin{matrix} 1 & 1 & 0 \\ 1&1&0  \\ 0&1&1 
	\end{matrix} & \cdots & 0  \\
	\vdots & \ddots & \vdots \\
	0 & \cdots & \begin{matrix}   1 & 1 & 0 \\ 1&1&0  \\ 0&1&1  \end{matrix}
\end{array}
\right)
$}
\end{tabular}
\caption{Two adjacency matrices,  $\wt{A}_{1}$,  and  $\wt{A}_{2}$ } \label{matrices}
\end{center}
\end{table} 

The corresponding random intersection graph  assigns edges between all agents which insure the same objects. 
So the graphs are very different, model 1 has one strongly connected component, whereas model 2 decomposes in small independent components. Intuitively they would also be very different concerning risk contagion. 
In both cases, however, the average agent degrees as well as the average object degrees are equal to $2$ -- indeed in Model 1 all degrees  equal 2.
In our framework we compare  univariate and multivariate risk measures, where the difference between univariate and multivariate is manifested in  the univariate and multivariate network constants  from \eqref{VaRconst}. 
We calculate that 
\begin{eqnarray*}
C^{i}_{ind}(\wt{A}_{1})&=&  2^{1 - \alpha }    \quad \text{and} \\
 C^{i}_{ind}(\wt{A}_{2})&=& \left\{
 \begin{array}{ll}
  2^{-\alpha} + 3^{-\alpha} \quad & \mbox{ if } i ({\rm mod} \, d) = 1,2 ,\\
   1 + 3^{-\alpha} \quad & \mbox{ if } i ({\rm mod} \, d) = 0,
 \end{array}\right.
\end{eqnarray*} 
and hence the individual risk constants differ.

\subsection{Diversification benefit}\label{s52}

{In this section we discuss the notion of diversification benefit in the networks setting, and its relationship to the connectivity in the network.}

It is well known that for i.i.d.\ random variables  $X_{1},\dots,X_{q}\in\calr(-\al)$ the Value--at--Risk is asymptotically subadditive if $\alpha>1$ and asymptotically superadditive if $\alpha<1$:
\beao
\lim_{\gamma \to 0} \frac{\VaR_{1-\gamma}(X_{1}+\dots+X_{q})}{\VaR_{1-\gamma}(X_{1})+\dots+\VaR_{1-\gamma}(X_{q})} &\leq& 1 \quad\mbox{if } \alpha>1,\\
\lim_{\gamma \to 0} \frac{\VaR_{1-\gamma}(X_{1}+\dots+X_{q})}{\VaR_{1-\gamma}(X_{1})+\dots+\VaR_{1-\gamma}(X_{q})} &\geq &1
\quad\mbox{if } \alpha<1,
\eeao
with equality (additivity) for $\al=1$ (cf. \cite{Dan2005, ELW, ENW2009, MR, RK99}).
Then 
\begin{gather}\label{divben}
D := 1-\lim_{\gamma \to 0}\frac{ \VaR_{1-\gamma}(X_{1}+\dots+X_{q})  }{\VaR_{1-\gamma}(X_{1})+\dots+\VaR_{1-\gamma}(X_{q})} 
\end{gather}
has been interpreted as a \textit{diversification benefit} (cf. \cite{cope2009challenges,ELW}). Clearly, in case of $D>0,$ the $\VaR$ is asymptotically subadditive, in case of $D<0$, the $\VaR$ is asymptotically superadditive, and $D=0$ is equivalent to asymptotical additivity of $\VaR.$
It is possible to calculate $D$ also for the Conditional Tail Expectation instead of the Value-at-Risk. 
By Lemma~\ref{Prop:ES} the expression for $D$  is exactly the same.

In the context of a multivariate regularly varying vector $F$, with dependence introduced by the random graph structure, we  now discuss the influence of this graph structure on 
$D$ and the question of additivity.
As we recognize in view of  Proposition~\ref{Cor:VaR}, insight about the effect of the network structure on  the Value-at-Risk and the Expected Shortfall, for the individual as well as the market risk,  is solely contained in the two constants $C^i_{ind}$ and $   C^S_{ind}$ given in \eqref{VaRconst}
where we are now concerned with the $1$-norm only.
Calculating the quantity \eqref{divben} using as notation ($U$ indicates the univariate setting and $S$ as before the systemic setting)
$$C^U_{ind} := \sum_{i=1}^q \left( C^i_{ind} \right)^{1/\alpha }=\sum_{i=1}^q\Big(\sum_{j=1}^{d}\E A_{ij}^{\alpha}\Big)^{1/\alpha} $$
 gives 
\begin{align}\label{divben1}
D=1- \lim_{\gamma \to 0}\frac{\VaR_{1-\gamma} (\|F\|_{1})  }{ \sum_{i=1}^{q}\VaR_{1-\gamma}(F_{i})  } = 1- \frac{ \left( C_{ind}^S\right)^{1/\alpha }}{ C_{ind} ^U}.
\end{align}
Consequently,  for a given tail index $\al$, $D$ depends not on the absolute size of the constants $C^U_{ind} $ and $ \left( C_{ind}^S\right)^{1/\alpha}  $, but on their ratio.
Unfortunately, there is neither an  \textit{ad hoc} interpretation of $C^U_{ind} $ or $C^S_{ind} $  in terms of the network structure available, nor are they  computable in a simple way, particularly when dimensions are high; but the Poisson approximation in Proposition~\ref{poissonprop} can be employed if the edge probabilities are not too large. 

For the toy model \eqref{1bbsquare} with $r=1$ we calculate 
$$D = 1-\frac{3^{1/\al}}{3(1 +b(2^{1-\alpha}-1)+b^{2}(2^{1-\alpha}-1)+  b^{3}( 3^{1-\alpha} + 1 - 2^{2-\alpha}))^{1/\al}}.$$
This quantity is plotted in Figure~\ref{divben_homotoy} as a function of $b$, which shows that it increases in $b$ for $\al<1$ and decreases in $b$ for $\al>1$, though being negative all the way. This means, that \VaR\ is superadditive, but gets additive for $b=1.$ 
The quantity $D$ for the Rasch model is depicted in Figure~\ref{divben_rasch}; the behaviour is qualitatively similar, but in these two examples $D$ does not reach 0.

\begin{figure}
\subfigure[Homogeneous model]{\includegraphics[width=0.49\textwidth,height=0.35\textwidth]{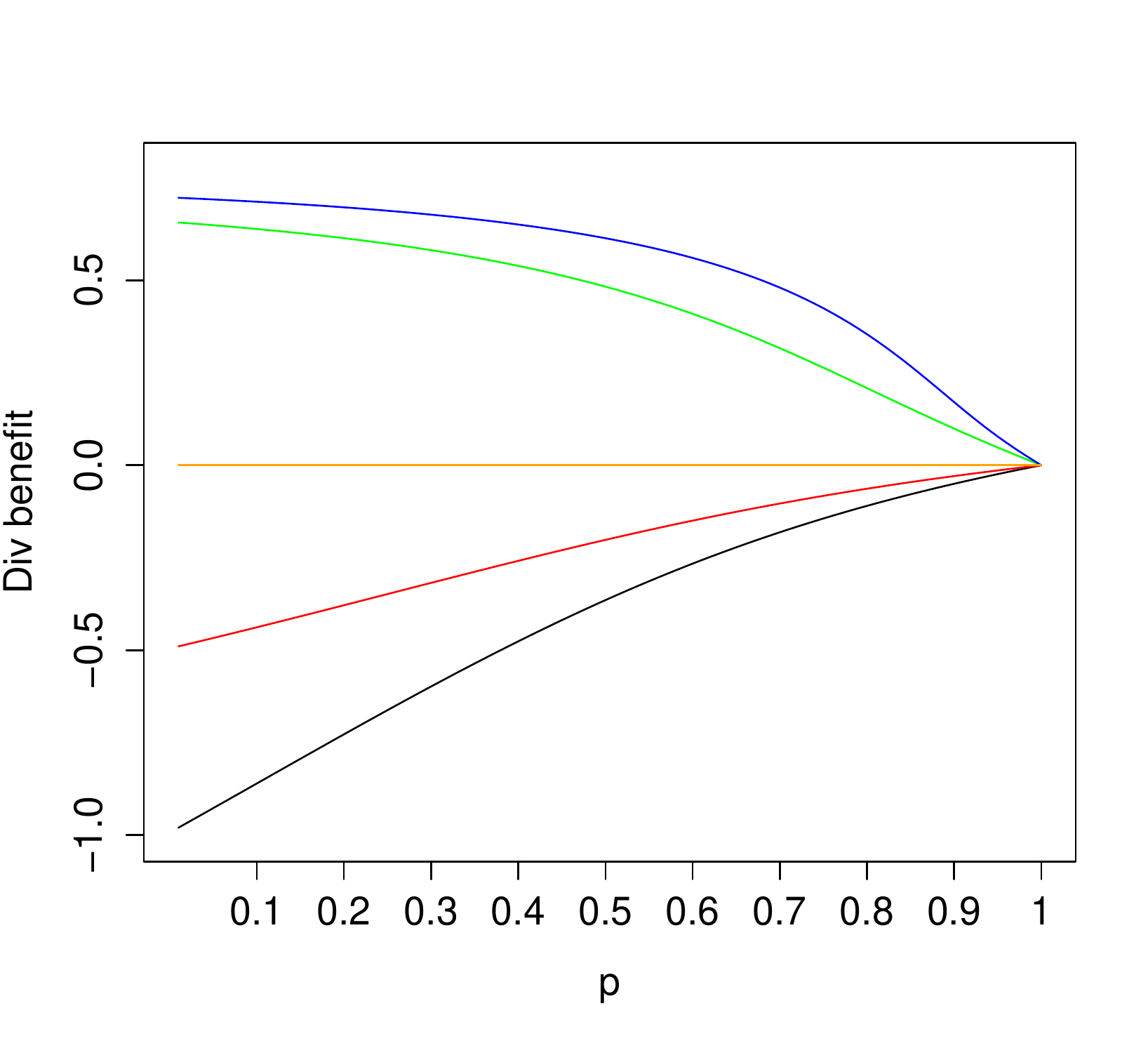}}\hfill
\subfigure[Toy model]{\includegraphics[width=0.49\textwidth,height=0.35\textwidth]{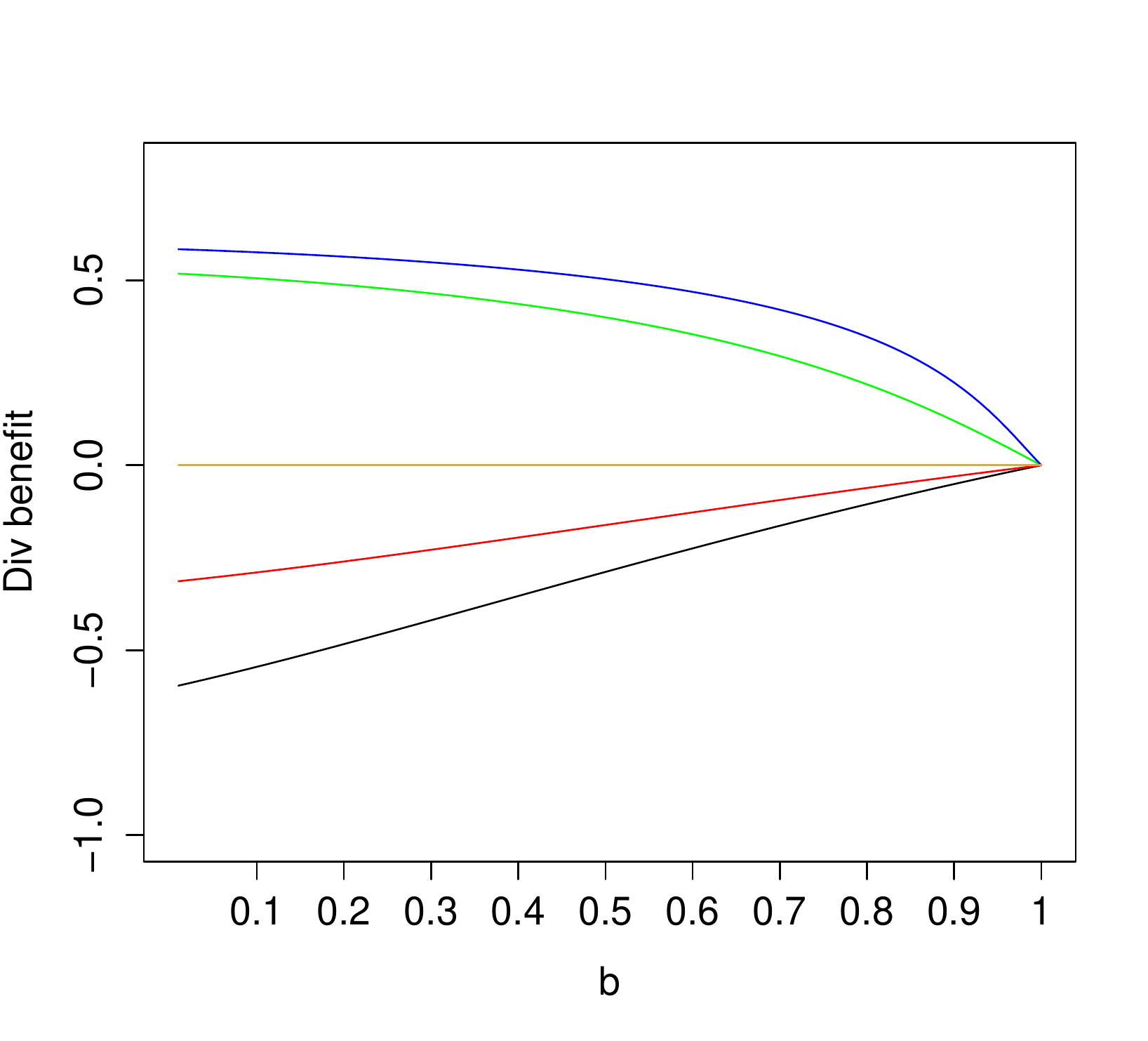}}
\caption{ \label{divben_homotoy}Diversification benefit for different tail indices $\alpha:$  orange for $\alpha=1$, blue for $\alpha= 5 $, green for $\alpha=3$ , red for $\alpha=0.8$ and  black for $\alpha=0.7$. }
\end{figure}

\begin{figure}
\vspace*{-1cm}
\subfigure[Fixed agent vector]{\includegraphics[width=0.49\textwidth,height=0.35\textwidth]{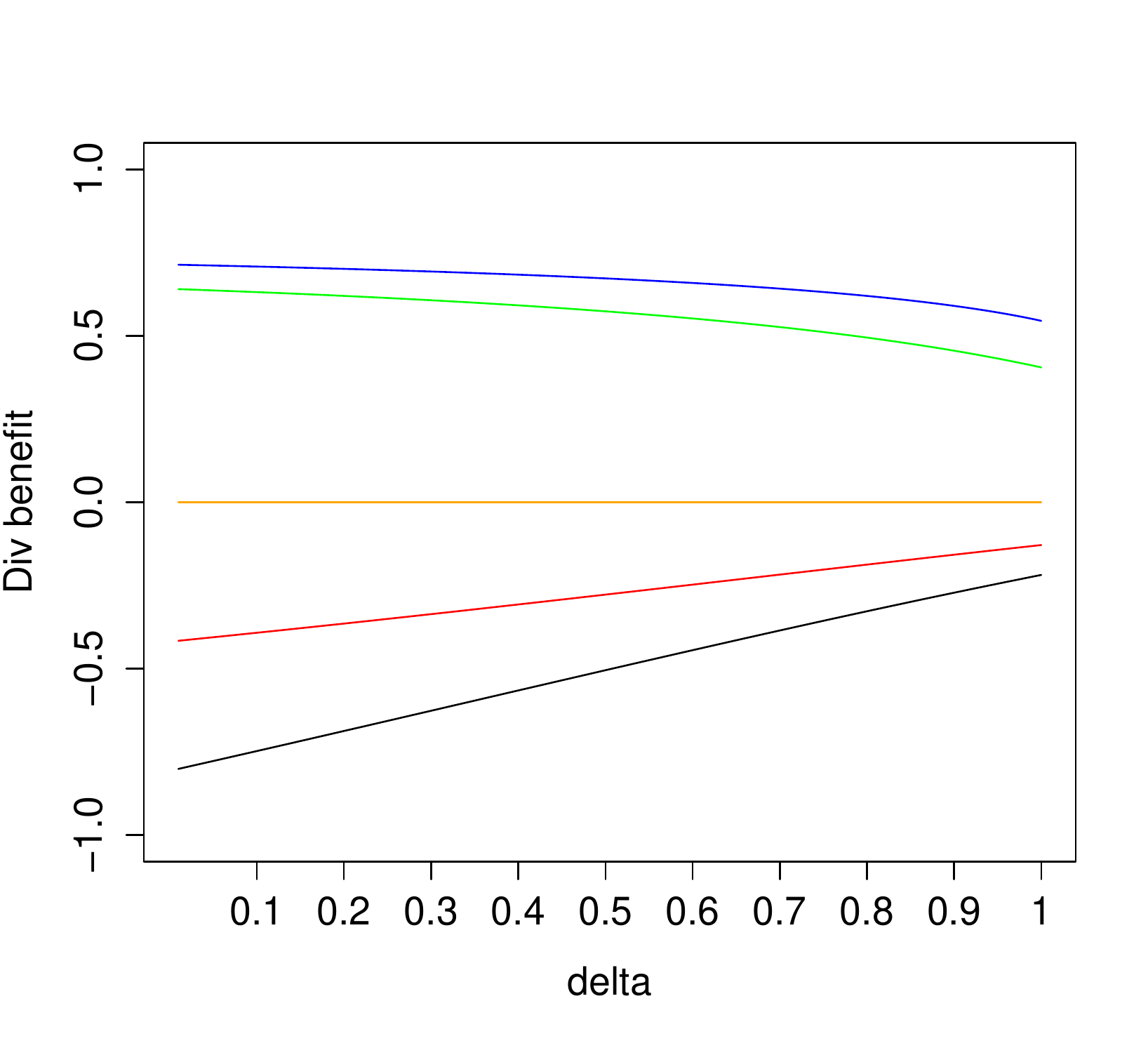}}\hfill
\subfigure[Fixed object vector]{\includegraphics[width=0.49\textwidth,height=0.35\textwidth]{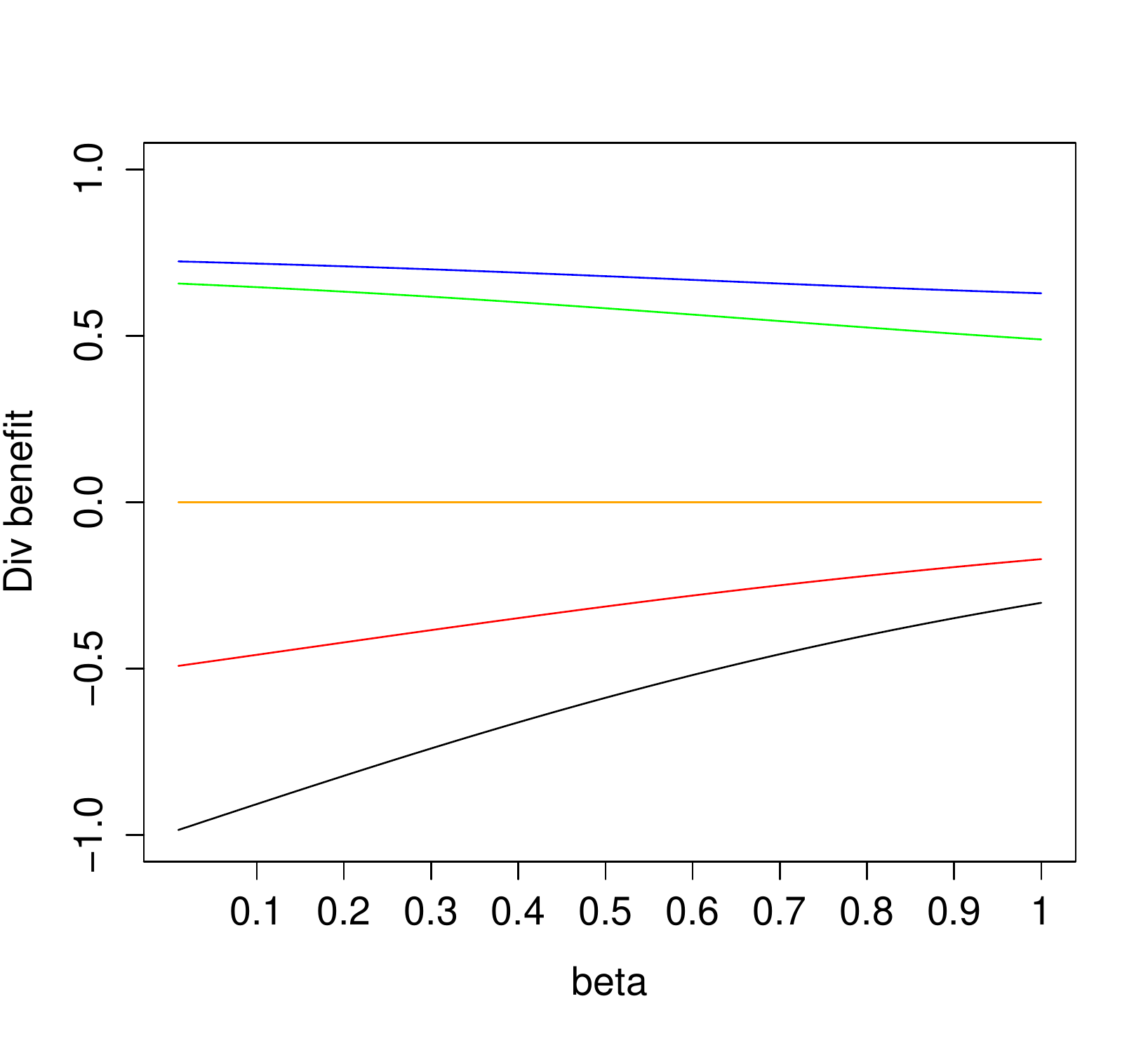}}
\caption{\label{divben_rasch} Rasch model: Diversification benefit for different tail indices $\alpha:$  orange for $\alpha=1$, blue for $\alpha= 5 $, green for $\alpha=3$ , red for $\alpha=0.8$ and  black for $\alpha=0.7$. The fixed vector is in both cases $(0.1,0.3,0.5,0.7,0.9)$}
\end{figure}

Whereas in Section~\ref{examples} we have interpreted risk by the constants \eqref{VaRconst} and \eqref{VaRconst_dep} (small constants means small risk),
we now focus on the definition of the diversification benefit in \eqref{divben1}. 
All of our examples share the feature that an increasing degree of connectivity in the network implies an increasing $D$ in the case of $\alpha<1$ and a decreasing $D$ in the case of $\alpha >1$. 
This is a simple consequence of the fact that the quantity in the numerator always decreases in $\alpha$, whereas the quantity in the denominator decreases for $\al>1$ and increases for $\al<1$. This implies that trivially $D$ increases for $\al<1$, whereas for $\al>1$ numerator and denominator go in opposite directions, leading in our examples to a weak decrease. 
In case of $\alpha=1,$ the diversification benefit equals 0 for all degrees of connectivity in the network.   If the degree of connectivity reaches its maximum; i.e., in the particular case of a complete bipartite graph, when each agent insures each object with probability one, $\VaR$ is asymptotically \textit{additive}. Indeed, 
\begin{gather}
\frac{ \left( C_{ind}^S\right)^{1/\alpha} }{ C_{ind} ^U}= \frac{\left( \sum_{j=1}^{d} \frac1{q^\al}\|(1,\dots,1)^{\top}\|_{1}^{\al}   \right)^{1/\alpha}}{\sum_{i=1}^{q}\left(\sum_{j=1}^{d} \frac1{q^\al}\right)^{1/\alpha} }=1.
\end{gather}

\subsection{\Gr{The issue} of $r$-norms and quasinorms} \label{other}

\Gr{In this final subsection we discuss the choice of the 1-norm and indicate alternative choices. This discussion is based on the axiomatic framework for systemic risk which is suggested in
\cite{axiomsystemic}, \cite{hannes} and \cite{overbeck}.   This general framework assumes that a  systemic risk measure $\rho$ of a multivariate risk $F = (F_1,\ldots, F_q)$  can be represented as the composition of a univariate (single-agent)  risk measure $\rho_0$ with an {\it{aggregation function}} $\Lambda: \mathbb{R}^q \rightarrow \mathbb{R}^q$, so that $\rho = \rho_0 \circ \Lambda$. Here, $\rho_0$ is usually assumed to be convex as well as monotone and positively 1-homogeneous; typical examples are the univariate Value-at-Risk and Contitional Tail Expectation measures. While detailed conditions on the aggregation function $\Lambda$ vary, there is consensus that  $\Lambda$ should satisfy $\Lambda(ax) = a \Lambda(x)$ for $a > 0$, and $\Lambda((1,\dots,1)^{\top})=q$. The last condition is satisfied for example for $\Lambda = \| \cdot \|_1 $ the 1-norm, but it is not satisfied for the norm or quasinorm $\| x \|_r=(\sum_{i=1}^{q} |x_{i}|^{r})^{1/r} $ for $0 < r<1$ or $r > 1$. 

Here we briefly discuss the cases $0 < r<1$ or $r > 1$. 
For monotone $\rho_0$,  we have $q<\|(1,\dots,1)\|_r$ for $0<r<1$ while $q>\|(1,\dots,1)\|_r$ for $1<r\leq \infty$. While the constants in $ C^i_{ind} $ and $ C_{ind}^S$ in  \eqref{VaRconst}, and $C_{dep}^i $ and $ C_{dep}^S$ in  \eqref{VaRconst_dep} are still defined, they may increase faster or increase slower, respectively, as the number of individual risks grows compared  to the constants when using the 1-norm as  aggregation function. 

Such effects  can be desired to account for the assumption that risk in a larger market may not increase strictly proportional to size, justifying using $r>1$. Similarly,   a small market in which the regulator may strive for more risk capital than the sum of risks, or the desire by a regulator to guard against 
 moral hazard from the different institutions, may 
suggest to use $r<1$.}

\OK{
However,  as indicated in Section~\ref{s5}, for $r>1$ such a risk measure allows for regulatory arbitrage! \GR{Any loss of size $V_j$ could be made smaller by splitting it between more agents even when the agents do not insure any losses, leading to a perverse interpretation of the risk measure.} Hence \Gr{$r > 1$} should be used with care, if at all.
Figure~\ref{multidiffnormsplots} presents the risk constants for different norms  in case of the homogeneous model. The fact that for $r>1$ risk decreases for increasing $p$ happens since the $r-$norms with $r>1$ favour diversification. \Gr{We stress that} \Ored{ the same mechanisms that allow for this diversification effect allow its interpretation as possibility for regulatory arbitrage.}

\begin{figure}  
\subfigure{\includegraphics[width=0.49\textwidth]{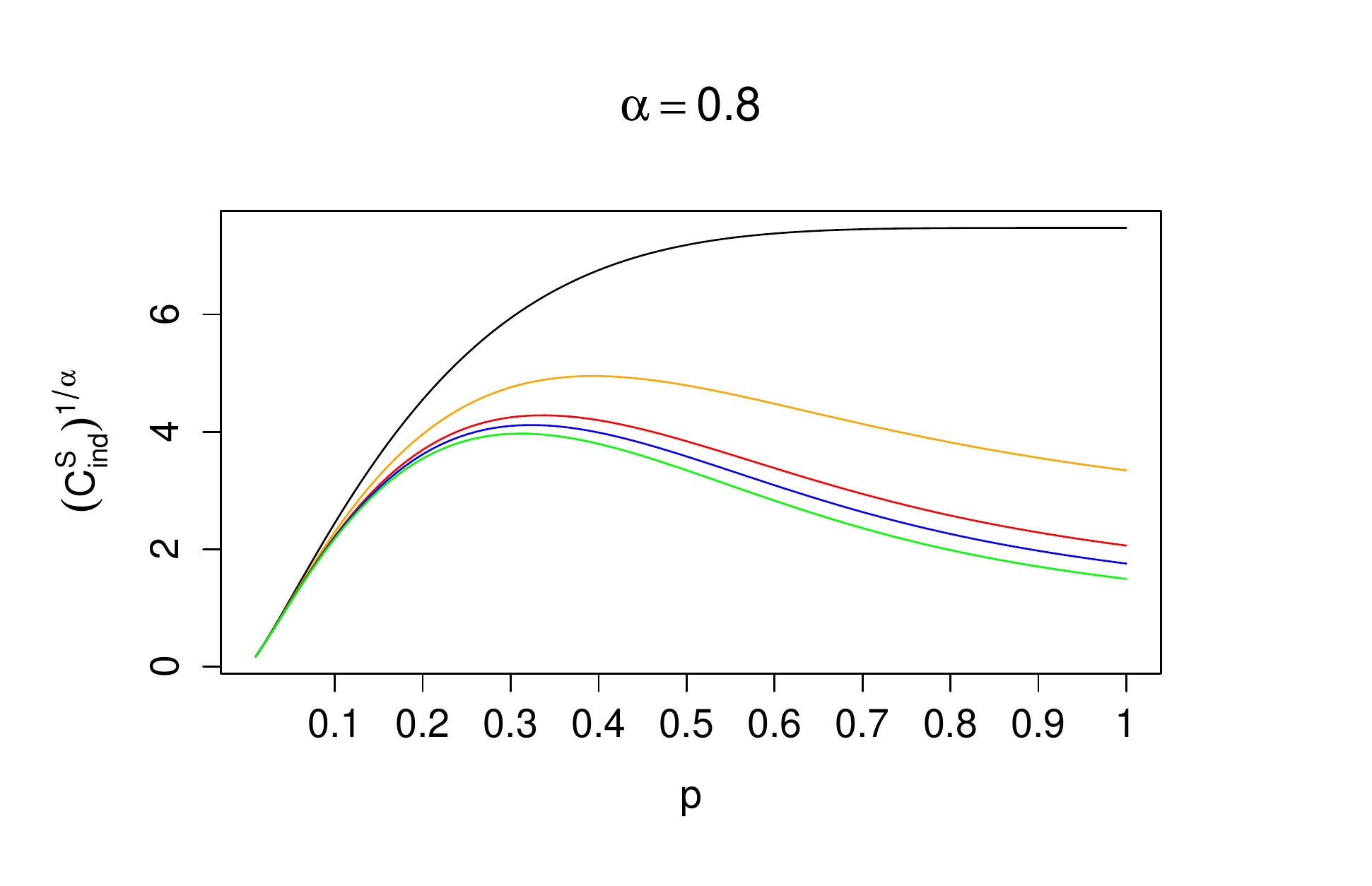}}\hfill
\subfigure{\includegraphics[width=0.49\textwidth]{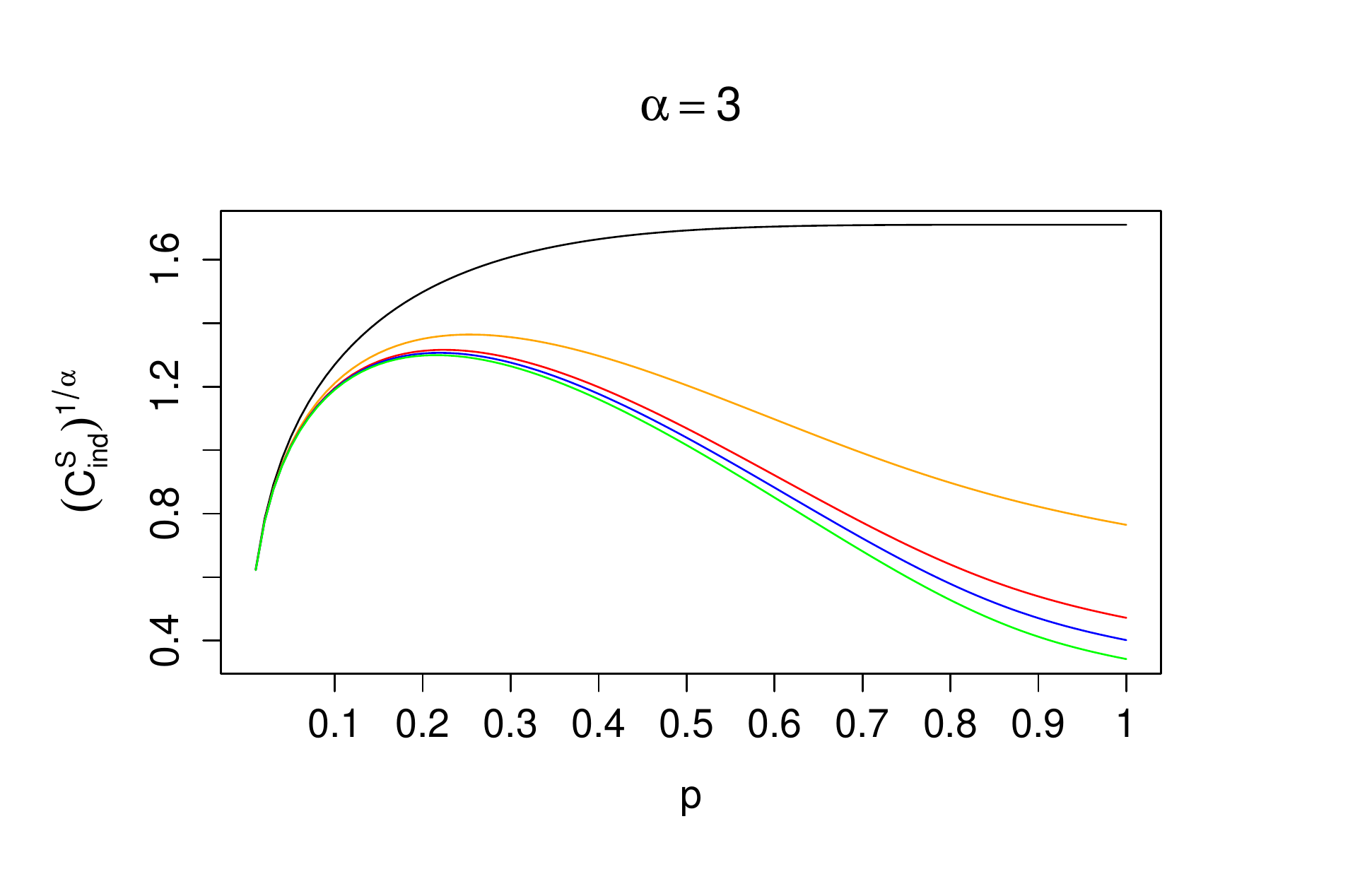}}
\caption{\label{multidiffnormsplots} Homogeneous probabilities: $(C^S_{ind})^{1/\al}$ for different norms: black for $r=1$, orange for $r=2$, red for $r=5$, blue for $r=10$, green for the max-norm}
\end{figure}

As a remedy we suggest to modify the weighted adjacency matrix $A$, incorporating such weights $W_{ij}$ in \eqref{eq2.2} to satisfy $\|A_{ij} V_j\|_r= V_j$. \Gr{Here we concentrate on the case $r>1$.} 
 Choosing $W_{ij} = \deg(j)^{1-1/r}$ gives 
\beam \label{newweight} \|A_{ij} V_j\|_r = \Big(\sum_{i=1}^q V_j^r \frac{\1(i\sim j)}{\deg(j)^{1-r}}\Big)^{1/r} =  V_j.\eeam 
In this setting, for $r>1$,  each agent's risk proportion is multiplied by $\deg(j)^{r}$.
This implies that the more agents insure an object, the more the risk is inflated.

This setting could serve as a model, when some agents default because of a large risks, then the surviving agents take over the defaulting agent's risk. 
Such types of contractual relationships are realistic and have similarly been 
implemented for example in the German Reinsurance Pharmapool or the German Nuclear Insurance Pool (see p.131  in \cite{MunichRe}).}

\Ored{For \eqref{newweight} \Gr{and $r > 1$} the individual constants take the form
 \begin{gather*}
 C^{i}_{ind}= \sum_{j=1}^d K_j \E\left( \deg(j)^{{1-1/r} }\1(i\sim j)\right)^{\alpha}\ \ \operatorname{and}\ \ 
 C^{i}_{dep}= \E\Big( \sum_{j=1}^d K_j^{1/\alpha}\deg(j)^{{1-1/r}} \1(i\sim j) \Big)^{\alpha}.
 \end{gather*}
}
\gr{Furthermore, 
\Ored{
$ || A e_j||_r = \big( \sum_{i=1}^q \1 ( i \sim j) (\deg(j))^{r-1} \big)^{1/r} =  \deg(j)$  
and, hence,
$$ C_{ind}^S 
= \sum_{j=1}^d K_j \deg(j)^{\alpha}\quad\mbox{and}\quad
C^S_{dep} =  \E \Big(\sum_{i=1}^q ( \sum_{j=1}^d K_j^{1/\al}  \deg(j)^{1-1/r}\1(i\sim j)   )^r \Big)^{\frac{\alpha}{r}}.
$$
}
\Gr{In contrast to \eqref{1constants_ind}, \eqref{1constants_S_ind} and \eqref{1constants_S_dep}, 
all four constants are now increasing in the  degree $\deg(j)$, and for the systemic constants the size of the degrees matter beyond whether or not they are 0. 
This observation is in contrast to our previous results and illustrates that the norm has to be chosen very carefully in order to reflect the features of interest of the system. 
}

%
%

}

\subsection*{Acknowledgements}

OK and CK gratefully acknowledge an invitation to the programme ``Systemic Risk: Mathematical Modelling and Interdisciplinary Approaches'' of the Isaac Newton Institute for Mathematical Sciences
in Cambridge, whose excellent working conditions promoted the project immensely. Our thanks also go to our colleagues attending the same programme and various discussions, which helped to clarify our thoughts. GR acknowledges support from EPSRC grant EP/K032402/1 as well as from the Oxford Martin School programme on Resource Stewardship. {{All authors would like to thank the referees for comments which have lead to a substantial improvement of the paper.}}

\bibliography{bibgesine}

\begin{thebibliography}{10}

\bibitem{CoVar}
T.~Adrian and M.K. Brunnermeier.
\newblock {CoVaR}.
\newblock Working Paper 17454, National Bureau of Economic Research, October
  2011.

\bibitem{AminiMinca}
H.~Amini and A.~Minca.
\newblock {Inhomogeneous Financial Networks and Contagious Links}.
\newblock Available at SSRN: http://ssrn.com/abstract=2518840 or
  http://dx.doi.org/10.2139/ssrn.2518840, 2014.

\bibitem{bollobas}
{B. Bollob{\'a}s}, C.~Borgs, J.~Chayes, and O.~Riordan.
\newblock {Directed scale-free graphs}.
\newblock In {\em {Proceedings of the Fourteenth Annual ACM-SIAM Symposium on
  Discrete Algorithms (Baltimore, 2003)}}, pages 132--139. ACM, New York, 2003.

\bibitem{bain1997insurance}
A.D. Bain.
\newblock {Insurance Spirals and the {L}ondon Market}.
\newblock {\em The Geneva Papers on Risk and Insurance - Issues and Practice},
  24(2):228--242, 1999.

\bibitem{Barbour_etal1992}
A.D. Barbour, L.~Holst, and S.~Janson.
\newblock {\em {Poisson Approximation}}.
\newblock Oxford University Press, Oxford, 1992.

\bibitem{BasrakPhD}
B.~Basrak.
\newblock {\em {The Sample Autocorrelation Function of Non-Linear Time
  Series}}.
\newblock PhD thesis, Rijksuniversteit Groningen, NL, 2000.

\bibitem{Basrak200295}
B.~Basrak, R.A. Davis, and T.~Mikosch.
\newblock {Regular variation of {GARCH} processes}.
\newblock {\em Stochastic Processes and their Applications}, 99(1):95--115,
  2002.

\bibitem{Biard}
R.~Biard.
\newblock {Asymptotic multivariate finite-time ruin probabilities with
  heavy-tailed claim amounts: impact of dependence and optimal reserve
  allocation}.
\newblock {\em Bulletin Francais d{\rq}Actuariat}, 13(26), 2013.

\bibitem{BGT1987}
N.H. Bingham, C.M. Goldie, and J.L. Teugels.
\newblock {\em {Regular Variation}}.
\newblock Cambridge University Press Cambridge ; New York, 1987.

\bibitem{BlanchetShi}
J.~Blanchet and Y.~Shi.
\newblock {Stochastic Risk Networks: Modeling, Analysis and Efficient {M}onte
  {C}arlo}.
\newblock 2012.

\bibitem{Bossetal}
M.~Boss, H.~Elsinger, M.~Summer, and S.Thurner.
\newblock {The network topology of the interbank market}.
\newblock {\em Quantitative Finance}, 4(6):677--684, 2004.

\bibitem{braverman2014networks}
A.~Braverman and A.~Minca.
\newblock {Networks of Common Asset Holdings: Aggregation and Measures of
  Vulnerability}.
\newblock {\em Available at SSRN 2379669}, 2014.

\bibitem{BKruin}
Y.~Bregman and C.~Kl{\"u}ppelberg.
\newblock {Ruin estimation in multivariate models with Clayton dependence
  structure.}
\newblock {\em Scand. Act. J.}, 2005(6):462--480, 2005.

\bibitem{Breiman}
L.~Breiman.
\newblock {On some limit theorems similar to the arc-sine law}.
\newblock {\em Theory Probab. Appl.}, 10:323--331, 1965.

\bibitem{Brownlees}
C.~T. Brownlees and R.~Engle.
\newblock {Volatility, correlation and tails for systemic risk measurement.},
  2010.
\newblock Working Paper Series, Department of Finance, NYU.

\bibitem{BrunnCher}
M.~K. Brunnermeier and P.~Cheridito.
\newblock {Measuring and Allocating Systemic Risk}.
\newblock Available at SSRN: http://ssrn.com/abstract=2372472 or
  http://dx.doi.org/10.2139/ssrn.2372472, 2014.

\bibitem{Cacciolietal}
F.~Caccioli, M.~Shrestha, C.~Moore, and J.D. Farmer.
\newblock {Stability analysis of financial contagion due to overlapping
  portfolios}.
\newblock SFI Working Paper: 2012-10-018, 2012.

\bibitem{axiomsystemic}
C.~Chen, G.~Iyengar, and C.C. Moallemi.
\newblock {An Axiomatic Approach to Systemic Risk}.
\newblock {\em Management Science}, 59(6):1373--1388, 2013.

\bibitem{MunichRe}
Munich~Reinsurance Company.
\newblock Annual report 2014.
\newblock
  \url{http://www.munichre.com/site/corporate/get/documents_E-1770937763/mr/assetpool.shared/Documents/0_Corporate%20Website/_Financial%20Reports/2015/Annual%20Report%202014/302-08574_en.pdf}.

\bibitem{CMS}
R.~Cont, A.~Moussa, and E.B. Santos.
\newblock {Network structure and systemic risk in banking systems}.
\newblock In J-P. Fouque and J.A. Langsam, editors, {\em {Handbook on Systemic
  Risk}}. Cambridge University Press, Cambridge, 2013.

\bibitem{cope2009challenges}
E.W. Cope, G.~Mignola, G.~Antonini, and R.~Ugoccioni.
\newblock {Challenges and pitfalls in measuring operational risk from loss
  data}.
\newblock {\em Journal of Operational Risk}, 4(4):3--27, 2009.

\bibitem{Dan2005}
C.G. de~Vries, G.~Samorodnitsky, B.N. Jorgensen, S.~Mandira, and J.~Danielsson.
\newblock {Subadditivity re--examined: the case for Value-at-Risk}.
\newblock FMG Discussion Papers dp549, Financial Markets Group, November 2005.

\bibitem{EK1}
I.~Eder and C.~Kl{\"u}ppelberg.
\newblock {The first passage event for sums of dependent {L}{\'e}vy processes
  with applications to insurance risk}.
\newblock {\em Ann. Appl. Probab.}, 19(6):2047--2079, 2009.

\bibitem{EK2}
I.~Eder and C.~Kl{\"u}ppelberg.
\newblock {Pareto {L}{\'e}vy measures and multivariate regular variation}.
\newblock {\em Advances in Applied Probability}, 44(1):117--138, 2012.

\bibitem{EN}
L.~Eisenberg and T.~Noe.
\newblock {Systemic risks in financial systems}.
\newblock {\em Management Science}, 47:2236--249, 2001.

\bibitem{ELW}
P.~Embrechts, D.D. Lambrigger, and M.V. W{\"u}thrich.
\newblock {Multivariate extremes and the aggregation of dependent risks:
  examples and counter-examples}.
\newblock {\em Extremes}, 12:107--127, 2009.

\bibitem{ENW2009}
P.~Embrechts, J.~Ne\v{s}lehov{\'a}, and M.V. W{\"u}thrich.
\newblock {Additivity properties for Value-at-Risk under Archimedean dependence
  and heavy-tailedness}.
\newblock {\em Insurance: Mathematics and Economics}, 44(2):164--169, 2009.

\bibitem{GaiKap}
P.~Gai and S.~Kapadia.
\newblock {Contagion in financial networks}.
\newblock {\em Proceedings of the Royal Society A: Mathematical, Physical and
  Engineering Science}, 466(2120):2401--2423, 2010.

\bibitem{haldane2011systemic}
A.G. Haldane and R.M. May.
\newblock {Systemic risk in banking ecosystems}.
\newblock {\em Nature}, 469(7330):351--355, 2011.

\bibitem{hannes}
H.~Hoffmann, T.~Meyer-Brandis, and G.~Svindland.
\newblock {Risk-Consistent Conditional Systemic Risk Measures}.
\newblock 2014.
\newblock preprint.

\bibitem{HLunpub}
H.~Hult and F.~Lindskog.
\newblock {Heavy-tailed insurance portfolios: buffer capital and ruin
  probabilities}.
\newblock Technical Report No. 1441, 2006.

\bibitem{IAIS}
International Association of Insurance~Supervisors (IAIS).
\newblock {Insurance and Financial Stability}, 2011.

\bibitem{Ibragimov2005}
R.~Ibragimov.
\newblock {Portfolio diversification and value at risk under thick-tailedness}.
\newblock {\em Quantitative Finance}, 9(5):565--580, 2009.

\bibitem{KK}
O.~Kley and C.~Kl{\"u}ppelberg.
\newblock {Bounds for randomly shared risk of heavy-tailed loss factors}.
\newblock arxiv:1503.03726[q-fin.RM], 2015.

\bibitem{KKR2}
O.~Kley, C.~Kl{\"u}ppelberg, and G.~Reinert.
\newblock {Conditional risk measures in a large claims market with bipartite
  graph structure}.
\newblock Preprint Technische Universit{\"a}t M{\"u}nchen, submitted, 2015.

\bibitem{overbeck}
E~Kromer, L.~Overbeck, and K.A. Zilch.
\newblock {Systemic Risk Measures on General Probability Spaces}.
\newblock 2014.

\bibitem{lin2014reinsurance}
Y.~Lin, J.~Yu, and M.O. Peterson.
\newblock {Reinsurance Networks and Their Impact on Reinsurance Decisions:
  Theory and Empirical Evidence}.
\newblock {\em Journal of Risk and Insurance}, 2014.

\bibitem{MR}
G.~Mainik and L.~R{\"u}schendorf.
\newblock {On optimal portfolio diversification with respect to extreme risks}.
\newblock {\em Finance and Stochastics}, 14(4):593--623, 2010.

\bibitem{rasch}
G.~Rasch.
\newblock {On general laws and the meaning of measurement in psychology}.
\newblock In {\em {Proceedings of the Fourth Berkeley Symposium on Mathematical
  Statistics and Probability, IV}}, pages 321--333. Berkeley, 1961.

\bibitem{Resnick1987}
S.I. Resnick.
\newblock {\em {Extreme Values, Regular Variation, and Point Processes}}.
\newblock Springer, New York, 1987.

\bibitem{Resnick2007}
S.I. Resnick.
\newblock {\em {Heavy-Tail Phenomena}}.
\newblock Springer, New York, 2007.

\bibitem{RK99}
H.~Rootz{\'e}n and C.~Kl{\"u}ppelberg.
\newblock {A single number can't hedge against economic catastrophes}.
\newblock {\em {AMBIO}: A Journal of the Human Environment}, 28(6), 1999.

\bibitem{SRTDWW}
G.~Samorodnitsky, S.~Resnick, D.~Towsley, R.~Davis, A.~Willis, and P.~Wan.
\newblock {Nonstandard regular variation of in-degree and out-degree in the
  preferential attachment model}.
\newblock arXiv:1405.4882v1 [math.PR], 2014.

\bibitem{GvPuninsured}
S.~von Dahlen and G.~von Peter.
\newblock {Natural catastrohpes and global reinsurance - exploring the
  linkages}.
\newblock {\em BIS Quarterly Review}, 2012.

\bibitem{Zhou}
C.~Zhou.
\newblock {Dependence structure of risk factors and diversification effects}.
\newblock {\em Insurance: Mathematics and Economics}, 46(3):531--540, 2010.

\bibitem{zigrand}
J.-P. Zigrand.
\newblock {Systems and Systemic Risk in Finance and Economics}, 2014.
\newblock SRC Special Paper No 1.

\end{thebibliography}
\bibliographystyle{plain}

\end{document}